\documentclass[prl,twocolumn,aps,showpacs,amssymb,superscriptaddress]{revtex4-1}

\usepackage{epsfig}
\usepackage{dcolumn}
\usepackage{amssymb}
\usepackage{amsmath}

\begin{document}

\title{\bf Temperature dependence of the pair coherence and healing lengths for a fermionic superfluid throughout the BCS-BEC crossover}

\author{F. Palestini}
\affiliation{Division of Physics, School of Science and Technology, Universit\`{a} di Camerino, 62032 Camerino (MC), Italy}

\author{G. C. Strinati}
\affiliation{Division of Physics, School of Science and Technology, Universit\`{a} di Camerino, 62032 Camerino (MC), Italy}
\affiliation{INFN, Sezione di Perugia, 06123 Perugia (PG), Italy}







\begin{abstract}
We calculate the pair correlation function and the order parameter correlation function, which probe, respectively, the intra-pair and inter-pair correlations of a Fermi gas with 
attractive inter-particle interaction, in terms of a diagrammatic approach as a function of coupling throughout the BCS-BEC crossover and of temperature, both in the superfluid and normal phase across the critical temperature $T_{c}$.
Several physical quantities are obtained from this calculation, including the pair coherence and healing lengths, the Tan's contact, the crossover temperature $T^{*}$ below which inter-pair correlations begin to build up in the normal phase, and the signature for the disappearance of the underlying Fermi surface which tends to survive in spite of pairing correlations.
A connection is also made with recent experimental data on the temperature dependence of the normal coherence length as extracted from the proximity effect measured in high-temperature (cuprate) superconductors.
\end{abstract}

\maketitle

\section{I. Introduction} 
\label{sec:introduction}

Pairing between fermions with opposite spins is at the essence of the theory of superconductivity \cite{Schrieffer-1964}.
In this context, Cooper pairs represent the building blocks on which macroscopic coherence is built up \cite{Cooper-2011}.
Roughly speaking, at zero temperature the wave function $\Psi(\boldsymbol{\rho},\mathbf{R}) \approx \phi(\boldsymbol{\rho}) \, \Phi(\mathbf{R})$ of a Cooper pair
contains information about the internal structure of the pair through $\phi(\boldsymbol{\rho})$ and about its center-of-mass motion through $\Phi(\mathbf{R})$,
where $\boldsymbol{\rho} = \mathbf{r} - \mathbf{r'}$ and $\mathbf{R} =  (\mathbf{r} + \mathbf{r'})/2$ are the relative and center-of-mass coordinates of the pair, in the order.

Two different lengths can be then associated with the spatial variations over the coordinates $\boldsymbol{\rho}$ and $\mathbf{R}$.
They are usually referred to as the \emph{pair coherence length} ($\xi_{\mathrm{pair}}$) and \emph{healing length} ($\xi_{\mathrm{phase}}$), and represent, respectively, 
the size of a Cooper pair and the spatial modulation of the pairs when subject to an external spatially varying disturbance \cite{footnote-xi-phase}.
Experimentally, the importance of the pair coherence length was first revealed by Pippard through the need for non-local electrodynamics, so that the pair coherence length
is sometimes referred to as the Pippard coherence length \cite{FW-1971}.
The healing length, on the other hand, has emerged from the Ginzburg-Landau differential equation and is associated with the spatial fluctuations of the superconducting order 
parameter (it sets, for instance, the spatial variation inside an Abrikosov vortex lattice close to the critical temperature \cite{Tinkham-1980}).
In both cases, a weak-coupling superconductor was considered for which Cooper pairs are strongly overlapping. 

With the advent of the BCS-BEC crossover \cite{KZ-2007} it became possible to follow the continuous evolution, from a situation where fermion (Cooper) pairs are strongly overlapping (which corresponds to the BCS weak-coupling limit), to a situation where fermion pairs form a dilute gas of composite bosons (which corresponds to the BEC strong-coupling limit).
At zero temperature, in the BCS limit one thus expects $\xi_{\mathrm{phase}}$ to coincide with $\xi_{\mathrm{pair}}$ (apart possibly from a multiplicative
factor of order unity due to their independent definitions), while in the BEC limit where the size of a pair shrinks to molecular dimensions one expects $\xi_{\mathrm{phase}}$ to be
much larger than $\xi_{\mathrm{pair}}$.
This expectation has been explicitly confirmed by numerical calculations done separately for $\xi_{\mathrm{pair}}$ \cite{PS-1994,MPS-1998} and $\xi_{\mathrm{phase}}$
\cite{PS-1996,Randeria-1997,MPS-1998,Randeria-2006}.
No calculation has, however, been performed to establish the temperature dependence of these two lengths throughout the BCS-BEC crossover, both below and above the critical temperature $T_{c}$ for the superfluid transition (with the exception of the work of Ref.\cite{Marsiglio-1990}, where $\xi_{\mathrm{pair}}$ was obtained in weak-coupling below $T_{c}$ within a BCS decoupling).
Purpose of the present paper is to fill this gap.

Albeit pictorially appealing, the wave function of a Cooper pair is strictly speaking an ill-defined concept, at least up to the point that the pairs become non-overlapping composite bosons when approaching the BEC limit of the crossover.
Quite generally, in the context of the many-body problem what can be addressed is the information about the \emph{intra-pair correlations} established between fermions of opposite spins and about the \emph{inter-pair correlations} relating different pairs.
The first one can be obtained from the pair correlation function $g_{\uparrow \downarrow}(\boldsymbol{\rho})$ that depends on the relative coordinate $\boldsymbol{\rho}$ of the pair, and the second one from the correlation function $\langle \Delta(\mathbf{R}) \, \Delta(\mathbf{R'}) \rangle$ of the order parameter $\Delta$ which depends on the difference $\mathbf{R} - \mathbf{R'}$ of the center-of-mass coordinates of different pairs (we consider a homogeneous system throughout).
We shall show that these correlation functions can be obtained, both below and above $T_{c}$, in terms of a common diagrammatic structure (which we shall keep at a minimal level to include the essential effects of pairing fluctuations), where only the variables at the end points of a common two-particle Green's function are set in different ways to identify the two functions. 

The key new physical results that we will obtain in this way can be summarized as follows:

\noindent
(i) The length $\xi_{\mathrm{pair}}$, which is obtained from $g_{\uparrow \downarrow}(\boldsymbol{\rho})$, is basically a decreasing function of temperature for given coupling $(k_{F} a_{F})^{-1}$ and remains finite at the corresponding value of $T_{c}$.
[Here, $a_{F}$ is the two-body scattering length and $k_{F}$ is the Fermi wave vector related to the density $n$ by $k_{F} = (3 \pi^{2} n)^{1/3}$ (in the following we shall consider a spin-balanced system).]
The rate of the decrease of $\xi_{\mathrm{pair}}$ turns out to be progressively less rapid when passing from the BCS to the BEC regimes.
This appears to be in line with a recent experimental finding for the normal coherence length $\xi_{\mathrm{N}}$ measured in the normal phase from the proximity effect occurring in a $\mathrm{SS'S}$ superconducting Josephson junction \cite{KK-2013}, once $\xi_{\mathrm{N}}$ is identified with $\xi_{\mathrm{pair}}$ above $T_{c}$ and to the extent that a stronger inter-particle coupling is attributed to the under-doped with respect to the optimally-doped regime of the high-temperature (cuprate) superconductor used in the experiment.

\noindent
(ii) The length $\xi_{\mathrm{phase}}$, which is obtained from the correlation function of the order parameter, is always larger than $\xi_{\mathrm{pair}}$ below $T_{c}$ for any coupling 
(provided the two independent definitions of $\xi_{\mathrm{phase}}$ and $\xi_{\mathrm{pair}}$ are suitably adjusted in the extreme BCS limit at zero temperature so as to have a single significant length in that limit \cite{PS-1996,PPS-2010}).
In addition, $\xi_{\mathrm{phase}}$ diverges at $T_{c}$ thus identifying the critical temperature.

\noindent
(iii) Above $T_{c}$, $\xi_{\mathrm{phase}}$ decreases more rapidly than $\xi_{\mathrm{pair}}$ for increasing temperature at a given coupling, such that a crossing of these two quantities is bound to occur at a certain temperature $T^{*}$.
On physical grounds, $T^{*}$ has then the meaning of a \emph{crossover temperature} below which independent pairs (whose partners are correlated over a finite length 
$\xi_{\mathrm{pair}}$) begin to build up an inter-pair correlation extending over the length $\xi_{\mathrm{phase}}$.
Precursor pairing phenomena (like, for instance, pseudo-gap effects \cite{Levin-2005}) are thus expected to occur only below $T^{*}$.

\noindent
(iv) Besides the length $\xi_{\mathrm{pair}}$, a detailed knowledge of the function $g_{\uparrow \downarrow}(\boldsymbol{\rho})$ provides also information about the underlying Fermi surface (if any) through its spatial oscillations.
This information can be related to the occurrence of a finite value of the Luttinger wave vector $k_{L}$, which can also be identified by the dispersion of the single-particle spectral function 
\cite{Camerino-Jila-2011}.

\noindent
(v) Interest in the pair correlation function $g_{\uparrow \downarrow}(\boldsymbol{\rho})$ has recently been revived in the context of the Tan's contact $C$, which is a measure of the number of fermion pairs in the two spin states at small separation and connects a number of universal relations involving the properties of a system with short-range dynamics \cite{Tan-2008,Braaten-2012}.
By the present approach, we correctly reproduce not only the leading limiting behavior $\displaystyle{\lim_{\boldsymbol{\rho}\rightarrow 0}} g_{\uparrow \downarrow}(\boldsymbol{\rho}) = C /(4 \pi \rho)^{2}$ where the coupling does not explicitly enter, but also the next sub-leading term $- C/(8 \pi^{2} a_{F} \rho)$ that contains the scattering length $a_{F}$ (here, $\rho = |\boldsymbol{\rho}|$).

\noindent
(vi) The pair correlation function is not a response function and thus is not bound to satisfy conservation criteria.
As a consequence, when using diagrammatic methods to calculate it, strictly speaking one cannot be guided by standard procedures of ``conserving approximations'' \cite{BK-1961,Baym-1962}.
In this context, we shall find it relevant to revive an argument given by Bell some time ago \cite{Bell-1963}, about a ``sum rule'' which should apparently be obeyed by $g_{\uparrow \downarrow}(\boldsymbol{\rho})$ once integrated over $\boldsymbol{\rho}$, but that in reality is satisfied in this sense only in the high-temperature (classical) limit.

For completeness, we mention that the internal structure of Cooper pairs at finite temperatures was also considered in Ref.\cite{Andrenacci-Beck-2006} through a diagrammatic pairing approach for the pair correlation function which bears some similarities to the present one.
However, in Ref.\cite{Andrenacci-Beck-2006} use was made of a different pairing theory (built on a quasi-two-dimensional single-band Hamiltonian in a lattice to make contacts with the physics of the cuprates) and calculations were limited to the spatial profile of $g_{\uparrow \downarrow}(\boldsymbol{\rho})$ at two specific values of the temperature in the normal phase, thus making essentially no reference to the physics of the BCS-BEC crossover with ultra-cold gases. 
None of the issues (i)-(vi) listed above were then discussed or even addressed in Ref.\cite{Andrenacci-Beck-2006}. 

The paper is organized as follows.
In Section II the pair correlation function for intra-pair correlations is obtained in terms of the many-body diagrammatic structure, and then explicitly calculated by going beyond the standard BCS approximation below $T_{c}$ such that a pairing approximation results correspondingly also above $T_{c}$.
Information about the pair coherence length $\xi_{\mathrm{pair}}$, the Luttinger wave vector $k_{L}$, and the Tan's contact $C$ is then extracted from the pair correlation function for all temperatures both below and above $T_{c}$ and for all couplings throughout the BCS-BEC crossover.
A comparison is also made of the temperature dependence of $\xi_{\mathrm{pair}}$ at various couplings with the available experimental data on the proximity effect in the normal phase of high-temperature (cuprate) superconductors.
In Section III the correlation function of the order parameter describing inter-pair correlations is obtained in terms of the same diagrammatic structure, and calculated again for all temperatures both below and above $T_{c}$ and for all couplings throughout the BCS-BEC crossover to obtain the healing length $\xi_{\mathrm{phase}}$.
The different temperature dependence resulting for $\xi_{\mathrm{pair}}$ and $\xi_{\mathrm{phase}}$ at given coupling is then exploited to identify a crossover temperature $T^{*}$ below which pairing effects are expected to become significant in physical quantities.
Section IV gives our conclusions.
Appendix A reconsiders an argument given originally by Bell at $T=0$ about the correct way to interpret a sum rule for $g_{\uparrow \downarrow}(\boldsymbol{\rho})$, and rephrases it into the terminology used in the present paper, in order to extend it to all temperatures and to check it numerically within the present approach.
Appendix B derives analytically the expressions of the asymptotic behavior of $\xi_{\mathrm{pair}}$ and $\xi_{\mathrm{phase}}$ at high temperature.
Appendix C discusses the relationship between $\xi_{\mathrm{N}}$ and $\xi_{\mathrm{pair}}$ or $\xi_{\mathrm{phase}}$.

\vspace{-0.2cm}
\section{II. The pair correlation function and the associated length $\xi_{\mathrm{pair}}$} 
\label{sec: xi-pair}
\vspace{-0.2cm}

In this Section, we calculate the spatial profile of the pair correlation function as a function of coupling and temperature in terms of a diagrammatic approach, from which
information can be obtained on several physical quantities that are of interest to the BCS-BEC crossover.

The physical system we are considering is a gas of fermions of mass $m$ with two equally populated spin components that mutually interact via a short-range attraction 
$v_{0} \, \delta(\mathbf{r} - \mathbf{r'})$ where $v_{0} < 0$.
In what follows, we regularize this interaction in terms of the scattering length $a_{F}$ in a standard way, by introducing an ultraviolet wave-vector cutoff $k_{0}$ such that
in the expression \cite{PS-2000} 
\begin{equation}
\frac{m}{4 \pi a_{F}} = \frac{1}{v_{0}} + \int^{k_{0}} \!\!\! \frac{d\mathbf{k}}{(2 \pi)^{3}} \frac{m}{\mathbf{k}^{2}}
\label{ultraviolet-regularization}
\end{equation}

\noindent
$k_{0} \rightarrow \infty$ and $v_{0} \rightarrow 0$ at the same time so as to keep $a_{F}$ at a desired value (we set $\hbar = 1$ throughout).

\vspace{0.1cm}
\begin{center}
{\bf A. General formalism}
\end{center}
\vspace{0.1cm}

Quite generally, the \emph{pair correlation function} for opposite-spin fermions is defined by:
\begin{small}
\begin{eqnarray}
g_{\uparrow \downarrow}(\boldsymbol{\rho})  \!\! & = & \!\! 
                             \left\langle \psi^{\dagger}_{\uparrow}\left(\mathbf{R}+\frac{\boldsymbol{\rho}}{2}\right) \! \psi^{\dagger}_{\downarrow}\left(\mathbf{R}-\frac{\boldsymbol{\rho}}{2}\right) \!
                                               \psi_{\downarrow}\left(\mathbf{R}-\frac{\boldsymbol{\rho}}{2}\right) \! \psi_{\uparrow}\left(\mathbf{R}+\frac{\boldsymbol{\rho}}{2}\right) \right\rangle
\nonumber \\
& - & \!\! \left( \frac{n}{2} \right)^{2}                                                                                     
\label{definition-pair-correlation-function}
\end{eqnarray}
\end{small}

\noindent
where $\psi_{\sigma}(\mathbf{r})$ is a fermion field operator with spin component $\sigma=(\uparrow,\downarrow)$ and $\langle \cdots \rangle$ is a thermal average.
In Eq.(\ref{definition-pair-correlation-function}) the dependence on the center-of-mass coordinate $\mathbf{R}$ drops out for the homogeneous system we are considering.

To deal with the superfluid and normal phases on the same footing, it is convenient to introduce at the outset the Nambu representation of the field operators, whereby 
$\Psi_{1}(\mathbf{r}) = \psi_{\uparrow}(\mathbf{r})$ and $\Psi_{2}(\mathbf{r}) = \psi^{\dagger}_{\downarrow}(\mathbf{r})$ with Nambu index $\ell=(1,2)$.
Introducing further the time ordering operator $T_{\tau}$ for imaginary time $\tau$, the expression (\ref{definition-pair-correlation-function}) can be rewritten in the form
\begin{eqnarray}
g_{\uparrow \downarrow}(\boldsymbol{\rho}) + \left( \frac{n}{2} \right)^{2} & =  & \left\langle T_{\tau}[ \Psi(1) \Psi(2) \Psi^{\dagger}(2') \Psi^{\dagger}(1') ] \right\rangle \nonumber \\
& = & \mathcal{G}_{2}(1,2;1',2')
\label{pair-correlation-function-Nambu}
\end{eqnarray}

\noindent
with the following compact notation for the variables:
\begin{eqnarray}
1 & = & (\boldsymbol{\rho}/2, \tau, \ell=1)           \nonumber \\
2 & = & (-\boldsymbol{\rho}/2, \tau^{++}, \ell=2)  \nonumber \\
1' & = & (-\boldsymbol{\rho}/2, \tau^{+}, \ell=2)   \nonumber \\
2' & = & (\boldsymbol{\rho}/2, \tau^{+++}, \ell=1)  
\label{compact-notation}
\end{eqnarray}

\noindent
where $\tau^{+}$ signifies that $\tau$ is augmented by a positive infinitesimal $\eta = 0^{+}$.

\begin{figure}[h]
\begin{center}
\includegraphics[angle=0,width=7.5cm]{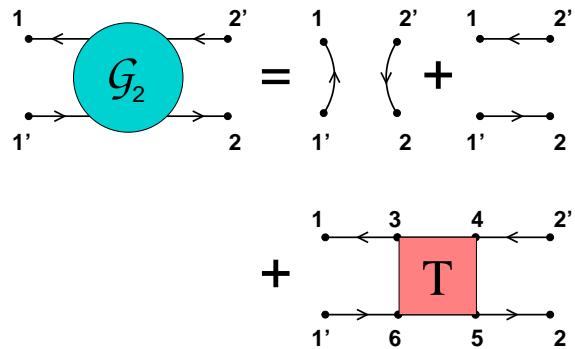}
\caption{(Color online) Diagrammatic representation of the two-particle Green's function $\mathcal{G}_{2}$ given by Eq.(\ref{Bethe-Salpeter-equation}), in terms of the single-particle Green's function 
              $\mathcal{G}$ and the many-particle T-matrix $T$. The arrows point from the second to the first argument of $\mathcal{G}$, and the variables stand for the set 
              $1=(\mathbf{r}_{1},\tau_{1},\ell_{1})$, and so on.}
\label{fig-1}
\end{center}
\end{figure}

Quite generally, the two-particle Green's function $\mathcal{G}_{2}$ in Eq.(\ref{pair-correlation-function-Nambu}) can be represented in terms of the single-particle Green's function $\mathcal{G}$ and the many-particle T-matrix $T$, in the form \cite{APS-2003}:
\begin{eqnarray}
& & \mathcal{G}_{2}(1,2;1',2') = \mathcal{G}(1,1') \mathcal{G}(2,2') - \mathcal{G}(1,2') \mathcal{G}(2,1') 
\nonumber \\
& - & \int \!\! d3456 \, \mathcal{G}(1,3) \mathcal{G}(6,1') T(3,5;6,4) \mathcal{G}(4,2') \mathcal{G}(2,5) 
\label{Bethe-Salpeter-equation}
\end{eqnarray}

\noindent
which is represented pictorially in Fig.÷\ref{fig-1}.

With the external variables given by Eq.(\ref{compact-notation}), the second term on the right-hand side of Eq.(\ref{Bethe-Salpeter-equation}) equals $(n/2)^{2}$ and thus cancels with the second term on the left-hand side of Eq.(\ref{pair-correlation-function-Nambu}).
At the same time, the first term on the right-hand side of Eq.(\ref{Bethe-Salpeter-equation}) equals $\mathcal{G}_{12}(\boldsymbol{\rho},\tau=0^{-})^{2}$ where $\mathcal{G}_{12}$ is the \emph{anomalous} single-particle Green's function which is non-vanishing only in the superfluid phase below $T_{c}$.
Interaction lines will appear explicitly in the last term on the right-hand side of Eq.(\ref{Bethe-Salpeter-equation}) through the many-particle T-matrix, whose presence is thus essential to get 
meaningful results for $g_{\uparrow \downarrow}(\boldsymbol{\rho})$ in the normal phase above $T_{c}$.

In the following, we shall discuss the use of different approximations for the calculation of $\mathcal{G}_{2}$, and thus of $g_{\uparrow \downarrow}(\boldsymbol{\rho})$ according to
Eq.(\ref{pair-correlation-function-Nambu}).

\vspace{0.1cm}
\begin{center}
{\bf B. Results within the BCS decoupling}
\end{center}
\vspace{0.1cm}

The simplest approximation for the pair correlation function below $T_{c}$ consists in retaining only the first term on the right-hand side of Eq.(\ref{Bethe-Salpeter-equation}) such that
$g_{\uparrow \downarrow}(\boldsymbol{\rho}) = \mathcal{G}_{12}(\boldsymbol{\rho},\tau=0^{-})^{2}$, and in further approximating $\mathcal{G}_{12}$ by its mean-field BCS expression
as follows:
\begin{eqnarray}
\mathcal{G}_{12}(\boldsymbol{\rho},\tau=0^{-}) & = & \int \! \frac{d\mathbf{k}}{(2 \pi)^{3}} e^{i \mathbf{k} \cdot \boldsymbol{\rho}} \, 
                                                                                     k_{B} T \sum_{n} e^{i \omega_{n} \eta} \mathcal{G}_{12}(\mathbf{k},\omega_{n})
\nonumber \\
& = & \Delta \, \int \! \frac{d\mathbf{k}}{(2 \pi)^{3}} e^{i \mathbf{k} \cdot \boldsymbol{\rho}} \, \frac{\left[ 1 - 2 f_{F}(E(\mathbf{k})) \right]}{2 E(\mathbf{k})} 
\label{G12-BCS}
\end{eqnarray}

\noindent
where $\mathcal{G}_{12}(\mathbf{k},\omega_{n}) = \Delta / (E(\mathbf{k})^{2} + \omega_{n}^{2})$.
Here, $k_{B}$ is the Boltzmann constant, $\omega_{n}=(2n+1) \pi k_{B} T$ ($n$ integer) a fermionic Matsubara frequency at temperature $T$, 
$\Delta$ the temperature-dependent BCS gap, $E(\mathbf{k}) = [\xi(\mathbf{k})^{2} + |\Delta|^{2}]^{1/2}$ with $\xi(\mathbf{k}) = \mathbf{k}^{2}/(2m) - \mu$ where $\mu$ is the chemical potential, 
and $f_{F}(\epsilon) = (\exp{(\epsilon/k_{B}T)} + 1)^{-1}$ is the Fermi function.

Accordingly, within the BCS decoupling we write for the volume integral of the distribution $g_{\uparrow \downarrow}(\boldsymbol{\rho})$ 
\begin{equation}
\int \! d \boldsymbol{\rho} \, g_{\uparrow \downarrow}(\boldsymbol{\rho}) = \frac{\Delta^{2}}{4} \int \! \frac{d\mathbf{k}}{(2 \pi)^{3}}  
\left[ \frac{1 - 2 f_{F}(E(\mathbf{k}))}{E(\mathbf{k})}\right]^{2} \, ,
\label{volume-integral-BCS}
\end{equation}

\noindent
and for its second moment 
\begin{equation}
\int \! d \boldsymbol{\rho} \, \boldsymbol{\rho}^{2} \, g_{\uparrow \downarrow}(\boldsymbol{\rho}) = \frac{\Delta^{2}}{4} \!\! \int \! \frac{d\mathbf{k}}{(2 \pi)^{3}}  
\left| \nabla_{\mathbf{k}} \left(\frac{[1 - 2 f_{F}(E(\mathbf{k}))]}{E(\mathbf{k})} \right) \right|^{2}
\label{second-moment-BCS}
\end{equation}

\noindent
(we have taken $\Delta$ to be real without loss of generality).
The pair coherence length $\xi_{\mathrm{pair}}$ is then obtained as follows in terms of the above two integrals:
\begin{equation}
\xi_{\mathrm{pair}}^{2} = \frac{\int \! d \boldsymbol{\rho} \, \boldsymbol{\rho}^{2} \, g_{\uparrow \downarrow}(\boldsymbol{\rho})}
                                               {\int \! d \boldsymbol{\rho} \, g_{\uparrow \downarrow}(\boldsymbol{\rho})} \, .
\label{xi-pair-definition}
\end{equation}

In particular, at zero temperature one may exploit the analytical results of Ref.\cite{MPS-1998}, to obtain for the volume integral of Eq.(\ref{volume-integral-BCS}) the expression:
\begin{equation}
\int \! d \boldsymbol{\rho} \, g_{\uparrow \downarrow}(\boldsymbol{\rho}) \,\, _{\overrightarrow{(T \rightarrow 0)}} \frac{(2 m \Delta_{0})^{3/2}}{16 \sqrt{2} \pi} \,
                                            \sqrt{ \frac{\mu}{\Delta_{0}} + \sqrt{1 + \left( \! \frac{\mu}{\Delta_{0}} \! \right)^{2}} }
\label{volume-integral-BCS-T=0}
\end{equation}

\noindent
where $\Delta_{0}=\Delta(T=0)$.
This coincides with the expression of the condensate fraction of a Fermi gas reported in Ref.\cite{SMP-2005}.

\begin{figure}[h]
\begin{center}
\includegraphics[angle=0,width=8.5cm]{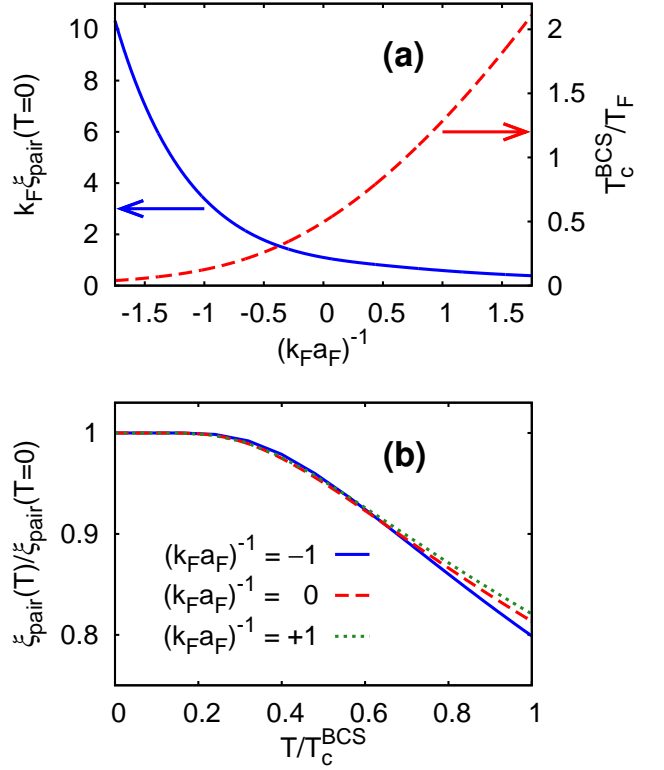}
\caption{(Color online) (a) BCS pair coherence length $\xi_{\mathrm{pair}}(T=0)$ at zero temperature in units of the inverse of the Fermi wave vector $k_{F}^{-1}$ (full line, left scale) and BCS 
               critical temperature $T^{\mathrm{BCS}}_{c}$ in units of the Fermi temperature $T_{F}=k_{F}^{2}/(2 m k_{B})$ (dashed line, right scale) vs the coupling $(k_{F} a_{F})^{-1}$.
              (b) BCS pair coherence length $\xi_{\mathrm{pair}}(T)$ in units of $\xi_{\mathrm{pair}}(T=0)$ vs the temperature $T$ in units of the respective BCS critical temperature $T^{\mathrm{BCS}}_{c}$ 
              at various couplings.}
\label{fig-2}
\end{center}
\end{figure}

The second moment of the distribution $g_{\uparrow \downarrow}(\boldsymbol{\rho})$ (and thus $\xi_{\mathrm{pair}}$) can also be obtained analytically for any coupling throughout the BCS-BEC crossover using the results Ref.\cite{MPS-1998}.
In particular, in the weak-coupling BCS limit (such that $(k_{F} a_{F})^{-1} \ll -1$) $\xi_{\mathrm{pair}}$ equals $1.11 \, \xi_{0}$ where $\xi_{0}=k_{F}/(\pi m \Delta_{0})$ is the Pippard coherence length, while in the strong-coupling BEC limit (such that $(k_{F} a_{F})^{-1} \gg +1$) $\xi_{\mathrm{pair}}$ reduces to the radius $a_{F}/\sqrt{2}$ of the two-body bound state \cite{PS-1994}.
For later reference, the coupling dependence of $\xi_{\mathrm{pair}}$ at $T=0$ is reported as the full curve of Fig.÷\ref{fig-2}(a).
Note that at unitarity (where $(k_{F} a_{F})^{-1} = 0$) the value of $k_{F} \xi_{\mathrm{pair}}(T=0)$ is approximately unity, meaning that the pair size is of the order of the inter-particle distance.

At finite temperature, $\xi_{\mathrm{pair}}$ can be obtained numerically from the expressions (\ref{volume-integral-BCS})-(\ref{xi-pair-definition}).
It turns out that the temperature dependence of $\xi_{\mathrm{pair}}$ is rather weak over the entire temperature interval from zero up to the BCS critical temperature $T^{\mathrm{BCS}}_{c}$.
This occurs not only in the weak-coupling regime (as already pointed out in Ref.\cite{Marsiglio-1990}) but also at stronger couplings, in such a way that $\xi_{\mathrm{pair}}$ always reaches a finite value 
at $T^{\mathrm{BCS}}_{c}$.
It turns further out that the temperature dependence of $\xi_{\mathrm{pair}}$ follows approximately a ``law of corresponding states'' irrespective of the coupling value, once 
$\xi_{\mathrm{pair}}(T)$ is expressed in units of $\xi_{\mathrm{pair}}(T=0)$ and $T$ is in units of $T^{\mathrm{BCS}}_{c}$.
This is shown in Fig.÷\ref{fig-2}(b) for a number of coupling values about unitarity, from which one sees that at the critical temperature $\xi_{\mathrm{pair}}$ decreases to only
about $80 \%$ of its value at $T=0$.
However, this kind of universal feature will not survive the inclusion of pairing fluctuations beyond mean field, to be considered in subsections II-C and II-D.

The above results have been obtained with the temperature-dependent values of $\Delta$ and $\mu$ which solve numerically the coupled BCS equations for the gap
\begin{equation}
- \frac{m}{4 \pi a_{F}} = \int \! \frac{d\mathbf{k}}{(2 \pi)^{3}} \left\{ \frac{\left[ 1 - 2 f_{F}(E(\mathbf{k})) \right]}{2 E(\mathbf{k})} - \frac{m}{\mathbf{k}^{2}} \right\} 
\label{BCS-gap-equation}
\end{equation}

\noindent
and for the density
\begin{equation}
n =  \int \! \frac{d\mathbf{k}}{(2 \pi)^{3}} \left\{ 1 - \frac{ \xi(\mathbf{k})}{E(\mathbf{k})} \left[ 1 - 2 f_{F}(E(\mathbf{k})) \right] \right\} \, .
\label{BCS-density-equation}
\end{equation}

\noindent
Recall in this context that the strong variation of $\mu$ when passing from the BCS to the BEC limits plays a crucial role in the physics of the BCS-BEC crossover \cite{KZ-2007}.

The finite value reached by $\xi_{\mathrm{pair}}$ upon approaching the critical temperature from below requires one to go beyond the simple BCS decoupling and include explicitly pairing
fluctuations even in the superfluid phase, as represented by the presence of the many-particle T-matrix in Fig.÷\ref{fig-1}.
Otherwise, when extrapolated to the normal phase, the simple BCS decoupling would yield $g_{\uparrow \downarrow}(\boldsymbol{\rho}) = 0$ and correspondingly $\xi_{\mathrm{pair}}$ could not be 
extracted from it.

An additional (and possibly more stringent) reason to include diagrams representing pairing fluctuations also in the superfluid phase stems from the short-range behavior of $g_{\uparrow \downarrow}(\boldsymbol{\rho})$
which is related to the Tan's contact \cite{Tan-2008,Braaten-2012}.
To see this, let's consider the $\boldsymbol{\rho} \rightarrow 0$ limit of the expression (\ref{G12-BCS}), which we manipulate as follows:
\begin{eqnarray}
& & \Delta \int \! \frac{d\mathbf{k}}{(2 \pi)^{3}} e^{i \mathbf{k} \cdot \boldsymbol{\rho}} \, \frac{\left[ 1 - 2 f_{F}(E(\mathbf{k})) \right]}{2 E(\mathbf{k})} =
\Delta \int \! \frac{d\mathbf{k}}{(2 \pi)^{3}} e^{i \mathbf{k} \cdot \boldsymbol{\rho}} \, \frac{m}{\mathbf{k}^{2}}
\nonumber \\
& + & \Delta \int \! \frac{d\mathbf{k}}{(2 \pi)^{3}} e^{i \mathbf{k} \cdot \boldsymbol{\rho}} \, \left\{ \frac{\left[ 1 - 2 f_{F}(E(\mathbf{k})) \right]}{2 E(\mathbf{k})} - \frac{m}{\mathbf{k}^{2}} \right\}
\,\,\,\, _{\overrightarrow{(\boldsymbol{\rho} \rightarrow 0)}}  
\nonumber \\
& = & \frac{m \, \Delta}{4 \pi \, \rho} + \Delta \int \! \frac{d\mathbf{k}}{(2 \pi)^{3}} \left\{ \frac{\left[ 1 - 2 f_{F}(E(\mathbf{k})) \right]}{2 E(\mathbf{k})} - \frac{m}{\mathbf{k}^{2}} \right\} + \cdots
\nonumber \\
& = & \left( \frac{m \Delta}{4 \pi} \right) \, \left( \frac{1}{\rho} - \frac{1}{a_{F}} + \cdots \right)
\label{contact-BCS-1}
\end{eqnarray}

\noindent
where in the last line we have made use of the BCS gap equation (\ref{BCS-gap-equation}).
The short-range behavior of $g_{\uparrow \downarrow}(\boldsymbol{\rho})$ which results from Eq.(\ref{contact-BCS-1}) is then given by:
\begin{equation}
g_{\uparrow \downarrow}(\boldsymbol{\rho})  \,\,  _{\overrightarrow{(\boldsymbol{\rho}\rightarrow 0)}} \,\, 
                                                                         \left( \frac{m \Delta}{4 \pi} \right)^{2} \, \left( \frac{1}{\rho^{2}} - \frac{2}{\rho \, a_{F}} + \cdots \right)
\label{contact-BCS-2}
\end{equation}

\noindent
where the factor $\left(m \Delta \right)^{2}$ represents the value of the Tan's contact $C$ within the BCS approximation.
Note that the dominant short-range behavior of $g_{\uparrow \downarrow}(\boldsymbol{\rho})$ in Eq.(\ref{contact-BCS-2}) stems from the ultraviolet behavior of the integral over the wave vector in Eq.(\ref{contact-BCS-1}).

The problem here is that, in the BCS limit, $\Delta$ (and thus $C$) is exponentially small in the coupling $(k_{F} a_{F})^{-1}$, while one would expect  on physical grounds the correct 
value of $C$ in this limit to be $(2 \pi a_{F} n)^{2}$, being it associated with a mean-field shift \cite{PPS-2009}.
Diagrams corresponding to pairing fluctuations over and above the BCS decoupling are therefore required to recover the expected value of $C$.
A related question is whether these additional diagrams may also somewhat modify the values of $\xi_{\mathrm{pair}}$ which was obtained within the BCS decoupling, as discussed next.

\vspace{0.1cm}
\begin{center}
{\bf C. Inclusion of pairing fluctuations below $T_{c}$}
\end{center}
\vspace{0.1cm}

We pass now to include the effect of the last term on the right-hand side of Eq.(\ref{Bethe-Salpeter-equation}) which contains the many-particle T-matrix.

\begin{figure}[h]
\begin{center}
\includegraphics[angle=0,width=9.0cm]{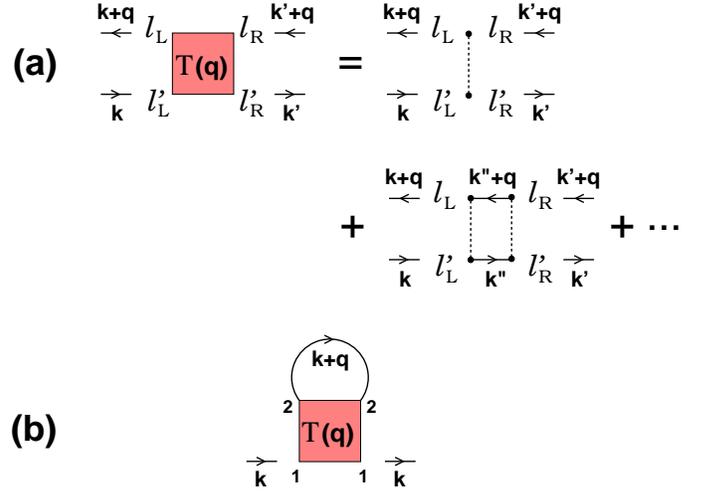}
\caption{(Color online) (a) Series of ladder diagrams for the T-matrix in the broken-symmetry phase. Conventions for four-momenta and Nambu indices are specified. Here, the dots delimiting the 
              potential (dashed lines) represent $\tau_{3}$ Pauli matrices. Only combinations with $\ell_{L} \ne \ell_{L}^{'}$ and $\ell_{R} \ne \ell_{R}^{'}$ occur owing to the regularization 
              (\ref{ultraviolet-regularization}) that we have adopted for the potential. (b) Fermionic self-energy diagram associated with the expression (\ref{self-energy-below-Tc}) below in the 
              broken-symmetry phase.}
\label{fig-3}
\end{center}
\end{figure}

Following Refs.\cite{APS-2003} and \cite{PPS-2004}, we approximate this term by \emph{the series of ladder diagrams} in the broken-symmetry phase which are depicted 
in Fig.÷\ref{fig-3}(a). 
Here, the interaction potential is taken to be of the short-range (contact) type and the lines represent the BCS single-particle Green's functions in Nambu notation: 
\begin{eqnarray}
\mathcal{G}_{11}(\mathbf{k},\omega_{n}) & = & - \mathcal{G}_{22}(-\mathbf{k},-\omega_{n}) = - \frac{\xi(\mathbf{k}) + i \omega_{n}}{E(\mathbf{k})^{2} + \omega_{n}^{2}}
\nonumber \\
\mathcal{G}_{12}(\mathbf{k},\omega_{n}) & = & \mathcal{G}_{21}(\mathbf{k},\omega_{n}) = \frac{\Delta}{E(\mathbf{k})^{2} + \omega_{n}^{2}}  \, .
\label{BCS-Green-functions}
\end{eqnarray}

\noindent
Making use of the convention $\mathrm{I} \leftrightarrow (\ell=1,\ell^{'}=2)$ and $\mathrm{II} \leftrightarrow (\ell=2,\ell^{'}=1)$ for the pairs of spin indices, the four independent elements of the T-matrix are then given by:
\begin{eqnarray}
\left( 
\begin{array}{cc}
-T_{\mathrm{I},\mathrm{I}}(q) & T_{\mathrm{I},\mathrm{II}}(q)  \\
T_{\mathrm{II},\mathrm{I}}(q)  & -T_{\mathrm{II},\mathrm{II}}(q)
\end{array}
\right) & = & \frac{1}{ A(q) A(-q) - B(q)^{2}}
\nonumber \\
& \times &
\left( 
\begin{array}{cc}
A(-q)  &  B(q)  \\
B(q)   &  A(q)
\end{array}
\right) \, .
\label{T-matrix-elements}
\end{eqnarray}

\noindent
In this expression, we have set $-A(q) = \frac{1}{v_{0}} + \Pi_{11}(q)$ and $B(q) =  \Pi_{12}(q)$, where
\begin{eqnarray}
\Pi_{11}(q) & = & \int \! dk \,\, \mathcal{G}_{11}(k+q) \, \mathcal{G}_{11}(-k)
\nonumber \\
\Pi_{12}(q) & = & \int \! dk \,\, \mathcal{G}_{12}(k+q) \, \mathcal{G}_{12}(-k)
\label{P_11_and_P_12}
\end{eqnarray}

\noindent
are particle-particle-like bubbles.
Here and in the following we adopt the four-vector notation $k=(\mathbf{k},\omega_{n})$ and $q=(\mathbf{q},\Omega_{\nu})$ ($\Omega_{\nu}=2 \pi k_{B} T \nu$ ($\nu$ integer) being a bosonic Matsubara frequency), and the short-hand notation
\begin{equation}
\int \! dk = \int \! \frac{d\mathbf{k}}{(2 \pi)^{3}} \, k_{B}T \sum_{n} 
\label{short-hand-notation}
\end{equation}

\noindent
with a similar expression for the four-integral over $q$.
Note that the wave-vector integral occurring in the definition of $\Pi_{11}(q)$ is ultraviolet divergent, in such a way that
\begin{equation}
R_{11}(q) \equiv \Pi_{11}(q) - \int \! \frac{d\mathbf{k}}{(2 \pi)^{3}} \, \frac{m}{\mathbf{k}^{2}} = -A(q) -\frac{m}{4 \pi a_{F}}
\label{R_11-vs-Pi_11}
\end{equation}

\noindent
is well behaved, where the regularization (\ref{ultraviolet-regularization}) has been utilized to obtain the last equality.

The series of ladder diagrams for the T-matrix depicted in Fig.÷\ref{fig-3}(a) is familiar in the context of gauge invariance in the response of a superconductor to an external electromagnetic field \cite{Schrieffer-1964}, which can quite generally be preserved provided the diagrammatic approximation one adopts is ``conserving'' in the sense of Baym and Kadanoff
\cite{BK-1961,Baym-1962}.
As we have already mentioned, however, no conservation law is associated with the pair correlation function (\ref{pair-correlation-function-Nambu}) of interest here, which can also be seen from the way the end variables (\ref{compact-notation}) are arranged in the two-particle Green's function where no time dependence appears (we shall return to this point more extensively in Appendix A).

When the above approximate form of the T-matrix is used in the expressions (\ref{pair-correlation-function-Nambu})-(\ref{Bethe-Salpeter-equation}), the pair correlation function below $T_{c}$ acquires the following form:
\begin{eqnarray}
& g_{\uparrow \downarrow}(\boldsymbol{\rho}) & \, = \, \mathcal{G}_{12}(\boldsymbol{\rho},\tau=0^{-})^{2} \! - \!\! \sum_{\ell_{3} \ell_{4} \ell_{5} \ell_{6}} \int \! dk dk' dq \, e^{i \Omega_{\nu} \eta}
e^{i (\mathbf{k}-\mathbf{k}') \cdot \boldsymbol{\rho} }
\nonumber \\
& \times & \!\! \mathcal{G}_{1 \ell_{3}}(k+q) \, \mathcal{G}_{\ell_{6} 2}(k) \, T^{\ell_{3} \ell_{4}}_{\ell_{6} \ell_{5}}(k,k';q) \, \mathcal{G}_{\ell_{4} 1}(k'+q) \, \mathcal{G}_{2 \ell_{5}}(k') 
\nonumber \\
& \longrightarrow & \,\, \mathcal{G}_{12}(\boldsymbol{\rho},\tau=0^{-})^{2} -  \int \! dk dk' dq \, e^{i \Omega_{\nu} \eta} e^{i (\mathbf{k}-\mathbf{k}') \cdot \boldsymbol{\rho} }
\nonumber \\
& \times & \left\{ \mathcal{G}_{11}(k+q) \mathcal{G}_{22}(k) \, T_{\mathrm{I},\mathrm{I}}(q) \, \mathcal{G}_{11}(k'+q) \mathcal{G}_{22}(k') \right.
\nonumber \\
& & + \, \mathcal{G}_{11}(k+q) \mathcal{G}_{22}(k) \, T_{\mathrm{I},\mathrm{II}}(q) \, \mathcal{G}_{21}(k'+q) \mathcal{G}_{21}(k') 
\nonumber \\
& & + \, \mathcal{G}_{12}(k+q) \mathcal{G}_{12}(k) \, T_{\mathrm{II},\mathrm{I}}(q) \, \mathcal{G}_{11}(k'+q) \mathcal{G}_{22}(k') 
\nonumber \\
& & \left. + \, \mathcal{G}_{12}(k+q) \mathcal{G}_{12}(k) \, T_{\mathrm{II},\mathrm{II}}(q) \, \mathcal{G}_{21}(k'+q) \mathcal{G}_{21}(k') \right\} 
\label{pair-correlation-function-below-Tc}
\end{eqnarray}

\noindent
where $\mathcal{G}_{12}(\boldsymbol{\rho},\tau=0^{-})$ is given by the mean-field expression (\ref{G12-BCS}).
This is the form of $g_{\uparrow \downarrow}(\boldsymbol{\rho})$ in terms of which we will calculate the effects of intra-pair correlations with the inclusion of pairing fluctuations below $T_{c}$.

In particular, the short-range behavior of $g_{\uparrow \downarrow}(\boldsymbol{\rho})$ which is contributed by pairing fluctuations can be obtained from the expression (\ref{pair-correlation-function-below-Tc}) in the following way.
Let's consider the factor that occurs in that expression:
\begin{eqnarray}
& & \int \! dk \, e^{i \mathbf{k} \cdot \boldsymbol{\rho}} \, \mathcal{G}_{11}(k+q) \mathcal{G}_{22}(k) 
\nonumber \\ 
& = & - \int \! dk \, e^{i \mathbf{k} \cdot \boldsymbol{\rho}} \, \mathcal{G}_{11}(k+q) \mathcal{G}_{11}(-k) = 
- \int \! \frac{d\mathbf{k}}{(2 \pi)^{3}} \, e^{i \mathbf{k} \cdot \boldsymbol{\rho}} \, \frac{m}{\mathbf{k}^{2}} 
\nonumber \\
& - & \int \! \frac{d\mathbf{k}}{(2 \pi)^{3}} \, e^{i \mathbf{k} \cdot \boldsymbol{\rho}} \, \left\{ k_{B}T \sum_{n} \mathcal{G}_{11}(k+q) \mathcal{G}_{11}(-k) - \frac{m}{\mathbf{k}^{2}} \right\} 
\nonumber \\
& & _{\overrightarrow{(\boldsymbol{\rho}\rightarrow 0)}} \,\,\, - \frac{m}{4 \pi \rho} - R_{11}(q)
\label{manipulation-1}
\end{eqnarray}

\noindent
with the notation of Eqs.(\ref{P_11_and_P_12}) and (\ref{R_11-vs-Pi_11}), since in the limit $\boldsymbol{\rho}\rightarrow 0$ we are allowed to set $e^{i \mathbf{k} \cdot \boldsymbol{\rho}} = 1$ in the last integral of Eq.(\ref{manipulation-1}) after it has been regularized.
Note that, here too, the dominant spatial short-range behavior stems from the ultraviolet behavior of the integral over the wave-vector.
By a similar token, we are allowed to set in the other factor occurring in the expression (\ref{pair-correlation-function-below-Tc}):
\begin{equation}
\int \! dk \, e^{i \mathbf{k} \cdot \boldsymbol{\rho}} \, \mathcal{G}_{12}(k+q) \mathcal{G}_{12}(k) \,\,\, _{\overrightarrow{(\boldsymbol{\rho}\rightarrow 0)}} \,\,\, \Pi_{12}(q)
\label{manipulation-2}
\end{equation}

\noindent
since this integral is convergent and does not require regularization.
Collecting the results (\ref{manipulation-1}) and (\ref{manipulation-2}), we then obtain for the short-range behavior of the fluctuation contribution in the expression (\ref{pair-correlation-function-below-Tc}):
\begin{eqnarray}
& & g_{\uparrow \downarrow}(\boldsymbol{\rho}) \, - \, \mathcal{G}_{12}(\boldsymbol{\rho},\tau=0^{-})^{2}  \,\,\,\,\,  _{\overrightarrow{(\boldsymbol{\rho}\rightarrow 0)}}
\nonumber \\
& - & \int \! dq \, e^{i \Omega_{\nu} \eta} \left\{ T_{\mathrm{I},\mathrm{I}}(q) \left( \frac{m}{4 \pi \rho} + R_{11}(q) \right)^{2} \right.
\nonumber \\
& - &  \left.  2 \, T_{\mathrm{I},\mathrm{II}}(q) \left( \frac{m}{4 \pi \rho} + R_{11}(q) \right) \, \Pi_{12}(q) + T_{\mathrm{II},\mathrm{II}}(q) \, \Pi_{12}(q)^{2} \right\}
\nonumber \\
& \simeq & - \left( \frac{m}{4 \pi \rho} \right)^{2} \,  \int \! dq \, e^{i \Omega_{\nu} \eta} \, T_{\mathrm{I},\mathrm{I}}(q)
\label{contact-fluctuation-1} \\
& - & 2 \, \left( \frac{m}{4 \pi \rho} \right) \,   \int \! dq \, e^{i \Omega_{\nu} \eta} \, \left\{ T_{\mathrm{I},\mathrm{I}}(q) \, R_{11}(q) - T_{\mathrm{I},\mathrm{II}}(q) \, \Pi_{12}(q) \right\} \, .
\nonumber
\end{eqnarray}

\noindent
Further manipulation of the last factor within braces in terms of the matrix elements (\ref{T-matrix-elements}) and of the relation (\ref{R_11-vs-Pi_11}) yields:
\begin{eqnarray}
& & T_{\mathrm{I},\mathrm{I}}(q) \, R_{11}(q) - T_{\mathrm{I},\mathrm{II}}(q) \, \Pi_{12}(q)
\label{manipulation-3} \\
& = & \frac{A(-q) \left( A(q) + \frac{m}{4 \pi a_{F}} \right) - B(q)^{2}}{ A(q) A(-q) - B(q)^{2}} = 1 \, - \,  \frac{m}{4 \pi a_{F}} \, T_{\mathrm{I},\mathrm{I}}(q) \, .
\nonumber
\end{eqnarray}

\noindent
In this way, Eq.(\ref{contact-fluctuation-1}) reduces to the simple result:
\begin{eqnarray}
& & g_{\uparrow \downarrow}(\boldsymbol{\rho}) \, - \, \mathcal{G}_{12}(\boldsymbol{\rho},\tau=0^{-})^{2}  \,\,\,\,\,  _{\overrightarrow{(\boldsymbol{\rho}\rightarrow 0)}}
\label{contact-fluctuation-2} \\
& - & \, \left( \frac{m}{4 \pi} \right)^{2} \, \int \! dq \, e^{i \Omega_{\nu} \eta} \, T_{\mathrm{I},\mathrm{I}}(q) \, \left( \frac{1}{\rho^{2}} - \frac{2}{a_{F} \, \rho} \, + \, \cdots \right) \, .
\nonumber 
\end{eqnarray}

\noindent
Following Ref.\cite{PPS-2009}, we then identify the pre-factor of Eq.(\ref{contact-fluctuation-2}) with the fluctuation contribution to the (square of the) high-energy scale $\Delta_{\infty}$, namely,
\begin{equation}
\Delta_{\infty}^{2} \, = \, - \, \int \! dq \, e^{i \Omega_{\nu} \eta} \, T_{\mathrm{I},\mathrm{I}}(q) \, ,
\label{Delta-infty-below}
\end{equation}

\noindent
such that $(m \, \Delta_{\infty})^{2}$ is the corresponding fluctuation contribution to the contact $C$.

Grouping together the mean-field contribution (\ref{contact-BCS-2}) and the fluctuation contribution (\ref{contact-fluctuation-2}), we obtain eventually for the short-range behavior of 
$g_{\uparrow \downarrow}(\boldsymbol{\rho})$ below $T_{c}$:
\begin{equation}
g_{\uparrow \downarrow}(\boldsymbol{\rho}) \,\,  _{\overrightarrow{(\boldsymbol{\rho}\rightarrow 0)}} \,\, \left( \frac{m}{4 \pi} \right)^{2} \left( \Delta^{2} + \Delta_{\infty}^{2} \right) \, 
                                                                                                                                                                                \left( \frac{1}{\rho^{2}} - \frac{2}{a_{F} \, \rho} \, + \, \cdots \right) 
\label{contact-BCS-plus-fluctuations}
\end{equation}

\noindent
where now the factor $m^{2} \left( \Delta^{2} + \Delta_{\infty}^{2} \right)$ is identified with the Tan's contact $C$ \cite{Haussmann-2009}.

\begin{figure}[h]
\begin{center}
\includegraphics[angle=0,width=7.3cm]{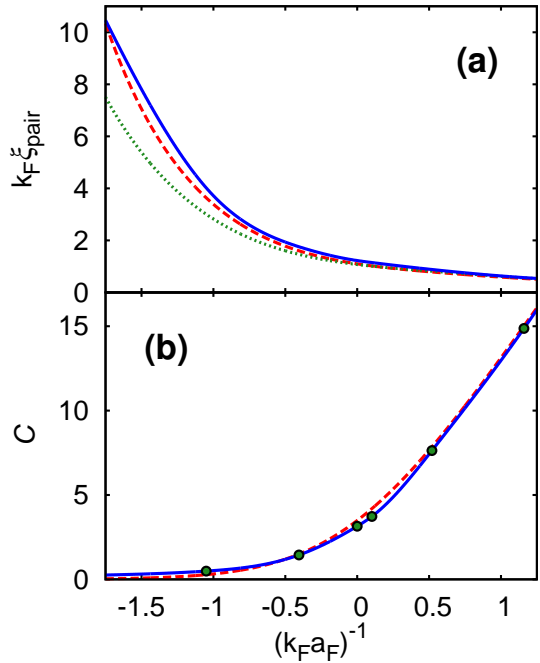}
\caption{(Color online) (a) Pair coherence length $\xi_{\mathrm{pair}}(T=0)$ at zero temperature in units of the inverse Fermi wave vector $k_{F}^{-1}$ vs the coupling $(k_{F} a_{F})^{-1}$, obtained 
               within mean field (dashed line) and with the inclusion of pairing fluctuations, where the gap and chemical potential are calculated either at the mean-field level according to 
               Eqs.(\ref{BCS-gap-equation}) and (\ref{BCS-density-equation}) (dotted line) or with the inclusion of fluctuations according to Eqs.(\ref{BCS-gap-equation}) and (\ref{modified-density-equation}) 
               (full line). (b) Coupling dependence of the Tan's contact $C$ at $T=0$, as obtained in terms of only the gap $\Delta$ at the mean-filed level (dashed line) and with the further inclusion of the 
               high-energy scale $\Delta_{\infty}$ given by Eq. (\ref{Delta-infty-below}) (full line). The dots represent the values of $C$ that were obtained in Ref.\cite{PPPS-2010} from the tail of the 
               wave-vector distribution $n(\mathbf{k})$.}
\label{fig-4}
\end{center}
\end{figure}

The character of universality which is intrinsic to the Tan's contact \cite{Tan-2008,Braaten-2012} implies that \emph{the same value} of $C$ that enters the pair-correlation function 
(\ref{contact-BCS-plus-fluctuations}) at short distances characterizes \emph{also} the tails of the wave-vector distribution $n(\mathbf{k})$ with 
$n = \int \! \frac{d\mathbf{k}}{(2 \pi)^{3}} n(\mathbf{k})$, such that $C = \displaystyle{\lim_{|\mathbf{k}| \rightarrow \infty}} |\mathbf{k}|^{4} n(\mathbf{k})$.
[An explicit comparison between the these two independent ways of obtaining the contact will be reported in Fig.÷\ref{fig-4}(b) below as a function of coupling at zero temperature.]
In the present context, the above argument implies that $n(\mathbf{k})$ cannot merely correspond to the expression within braces in the BCS density equation (\ref{BCS-density-equation}),
but should necessarily contain also the contribution of pairing fluctuations below $T_{c}$.

Accordingly, we are led to replace the BCS density equation (\ref{BCS-density-equation}) with the modified density equation discussed in Ref.\cite{PPS-2004}, whereby
\begin{equation}
n = 2 \int \! \frac{d\mathbf{k}}{(2 \pi)^{3}} \, k_{B}T \sum_{n} \, e^{i \omega_{n} \eta} \, G_{11}(\mathbf{k},\omega_{n})
\label{modified-density-equation}
\end{equation}

\noindent
with
\begin{equation}
G_{11}(\mathbf{k},\omega_{n}) = \frac{1}{i \omega_{n} - \xi(\mathbf{k}) - \sigma_{11}(\mathbf{k},\omega_{n})}
\label{modified-single-particle-Green-function}
\end{equation}

\noindent
and
\begin{equation}
\sigma_{11}(\mathbf{k},\omega_{n}) = \Sigma_{11}(\mathbf{k},\omega_{n}) + \frac{ \Delta^{2} }{i \omega_{n} + \xi(\mathbf{k}) + \Sigma_{11}(-\mathbf{k},-\omega_{n})} \, .
\label{effective-self-energy}
\end{equation}

\noindent
In the above expression, the self-energy $\Sigma_{11}$ is given by
\begin{equation}
\Sigma_{11}(k) = \int \! dq \,\, T_{\mathrm{I},\mathrm{I}}(q) \, \mathcal{G}_{11}(q-k) 
\label{self-energy-below-Tc}
\end{equation}

\noindent
with the short-hand-notation (\ref{short-hand-notation}), $T_{\mathrm{I},\mathrm{I}}$ given by Eqs.(\ref{T-matrix-elements}) and (\ref{P_11_and_P_12}), and $\mathcal{G}_{11}$ still of the BCS form (\ref{BCS-Green-functions}).
The self-energy (\ref{self-energy-below-Tc}) is represented diagrammatically in Fig.÷\ref{fig-3}(b).
On the other hand, according to Ref.\cite{PPS-2004} the gap equation maintains the BCS form (\ref{BCS-gap-equation}). 
As a consequence, new pairs of values for $\Delta$ and $\mu$ are obtained by solving Eqs.(\ref{modified-density-equation}) and (\ref{BCS-gap-equation}) for given coupling and temperature below $T_{c}$, 
values which have to be inserted into the expression (\ref{pair-correlation-function-below-Tc}) for $g_{\uparrow \downarrow}(\boldsymbol{\rho})$.

Correspondingly, the pair coherence length $\xi_{\mathrm{pair}}$ is obtained by entering the expression (\ref{pair-correlation-function-below-Tc}) into the definition (\ref{xi-pair-definition}).
Besides the mean-field contributions (\ref{volume-integral-BCS}) to the volume integral and (\ref{second-moment-BCS}) to the second moment of the distribution 
$g_{\uparrow \downarrow}(\boldsymbol{\rho})$, the fluctuation terms in the expression (\ref{pair-correlation-function-below-Tc}) contribute to these quantities as follows.
Let 
\begin{eqnarray}
\tilde{\Pi}_{11}(\mathbf{k};q) & \equiv & k_{B}T \sum_{n} \, \mathcal{G}_{11}(k+q) \, \mathcal{G}_{11}(-k)
\nonumber \\
\tilde{\Pi}_{12}(\mathbf{k};q) & \equiv & k_{B}T \sum_{n} \, \mathcal{G}_{12}(k+q) \, \mathcal{G}_{12}(-k)
\label{tildeP_11_and_tildeP_12}
\end{eqnarray}

\noindent
such that
\begin{eqnarray}
\Pi_{11}(q) & = &  \int \! \frac{d\mathbf{k}}{(2 \pi)^{3}} \, \tilde{\Pi}_{11}(\mathbf{k};q)
\nonumber \\
\Pi_{12}(q) & = &  \int \! \frac{d\mathbf{k}}{(2 \pi)^{3}} \, \tilde{\Pi}_{12}(\mathbf{k};q) \, .
\label{integralP_11_and_integralP_12}
\end{eqnarray}

\noindent
The fluctuation part $\delta g_{\uparrow \downarrow}(\boldsymbol{\rho}) \equiv g_{\uparrow \downarrow}(\boldsymbol{\rho}) - \mathcal{G}_{12}(\boldsymbol{\rho},\tau=0^{-})^{2}$
of the pair correlation function then contributes the terms
\begin{eqnarray}
& & \int \! d \boldsymbol{\rho} \, \delta g_{\uparrow \downarrow}(\boldsymbol{\rho}) = - \int \! dq \, e^{i \Omega_{\nu} \eta} \,T_{\mathrm{I},\mathrm{I}}(q) 
\! \int \! \frac{d\mathbf{k}}{(2 \pi)^{3}} \tilde{\Pi}_{11}(\mathbf{k};q)^{2}
\nonumber \\
& + & 2 \! \int \! dq \, e^{i \Omega_{\nu} \eta} \,T_{\mathrm{I},\mathrm{II}}(q) \! \int \! \frac{d\mathbf{k}}{(2 \pi)^{3}} \tilde{\Pi}_{11}(\mathbf{k};q) \, \tilde{\Pi}_{12}(\mathbf{k};q)
\nonumber \\
& - & \int \! dq \, e^{i \Omega_{\nu} \eta} \,T_{\mathrm{II},\mathrm{II}}(q) \! \int \! \frac{d\mathbf{k}}{(2 \pi)^{3}} \tilde{\Pi}_{12}(\mathbf{k};q)^{2}
\label{fluctuation-contribution-volume-integral}
\end{eqnarray}

\noindent
to the volume integral of $g_{\uparrow \downarrow}(\boldsymbol{\rho})$, and the terms
\begin{eqnarray}
& \int \! d \boldsymbol{\rho} & \boldsymbol{\rho}^{2} \delta g_{\uparrow \downarrow}(\boldsymbol{\rho}) = - \!\! \int \! dq e^{i \Omega_{\nu} \eta} T_{\mathrm{I},\mathrm{I}}(q) \!\! 
\int \! \frac{d\mathbf{k}}{(2 \pi)^{3}} \! \left[ \nabla_{\mathbf{k}} \tilde{\Pi}_{11}(\mathbf{k};q) \right]^{2}
\nonumber \\
& + & \!\!\! 2 \!\! \int \! dq \, e^{i \Omega_{\nu} \eta} T_{\mathrm{I},\mathrm{II}}(q) \!\! \int \! \frac{d\mathbf{k}}{(2 \pi)^{3}} \! \left[ \nabla_{\mathbf{k}} \tilde{\Pi}_{11}(\mathbf{k};q) \cdot \nabla_{\mathbf{k}} \tilde{\Pi}_{12}(\mathbf{k};q) \right]
\nonumber \\
& - & \int \! dq \, e^{i \Omega_{\nu} \eta} \,T_{\mathrm{II},\mathrm{II}}(q) \! \int \! \frac{d\mathbf{k}}{(2 \pi)^{3}} \left[ \nabla_{\mathbf{k}} \tilde{\Pi}_{12}(\mathbf{k};q) \right]^{2}
\label{fluctuation-contribution-second-moment}
\end{eqnarray}

\noindent
to its second moment. 

The results of this calculation are shown in Fig.÷\ref{fig-4}(a) where the coupling dependence of $\xi_{\mathrm{pair}}$ at zero temperature is reported, without (dashed line) and with (dotted and full lines) the inclusion of pairing fluctuations on top of mean field.
Here, the inclusion of pairing fluctuations further distinguishes the cases when $\Delta$ and $\mu$ are calculated either at the mean-field level (dotted line) or with the further inclusion of pairing fluctuations
(full line).
Note that in the first case (dotted line) there is a decrease of $\xi_{\mathrm{pair}}$ with respect to the mean-field value (dashed line).
Physically, this corresponds to the fact that the inclusion of quantum fluctuations 
at zero temperature over and above mean field tends to reduce the spatial extent over which correlations are effective.
On the other hand, when the values of $\Delta$ and $\mu$ are also affected by pairing fluctuations, the value of $\xi_{\mathrm{pair}}$ (full line) exceeds that at the mean-field level (dashed line),
because this procedure in practice has the effect of renormalizing the coupling to a smaller value.
Note also that, on the scale of Fig.÷\ref{fig-4}(a), the complete inclusion of pairing fluctuations modifies the mean-field result for $\xi_{\mathrm{pair}}$ only marginally.

The corresponding coupling dependence of the contact $C$ is shown in Fig.÷\ref{fig-4}(b).
Here, also reported for comparison are the values of $C$ obtained in Ref.\cite{PPPS-2010} from the tail of the integrand $n(\mathbf{k})$ of Eq.(\ref{modified-density-equation}) with the inclusion of fluctuations,
which show explicitly the character of universality associated with the Tan's contact.
[Note that $C$ is dimensionless provided the wave vectors are in units of $k_{F}$ ($n(\mathbf{k})$ is also normalized such that $ \int \! \frac{d\mathbf{k}}{(2 \pi)^{3}} \, n(\mathbf{k}) = \frac{1}{2}$).]

In this context, it is interesting to mention that the inclusion of pairing fluctuations in the broken-symmetry phase at low temperature is of interest also for problems in nuclear physics, where RPA calculations beyond BCS mean field are routinely performed \cite{Broglia-2005}.
 
\vspace{0.1cm}
\begin{center}
{\bf D. Pairing fluctuations above $T_{c}$}
\end{center}
\vspace{0.1cm}

Above $T_{c}$ where the gap $\Delta$ vanishes, only the first term within braces on the right-hand side of Eq.(\ref{pair-correlation-function-below-Tc}) survives.
In this case, $\mathcal{G}_{11}(k)$ reduces to the bare single-particle Green's function $G_{0}(k) = (i \omega_{n} - \xi(\mathbf{k}))^{-1}$ and $- T_{\mathrm{I},\mathrm{I}}(q)$ 
to the pair propagator $\Gamma_{0}(q) = - [1/v_{0} + \Pi_{0}(q)]^{-1}$ where
\begin{equation}
\Pi_{0}(q) = \int \! dk \, G_{0}(k+q) \, G_{0}(-k)
\label{particle-propagator-above-Tc}
\end{equation}

\noindent
is the particle-particle bubble. 
By defining further, in analogy to Eq.(\ref{tildeP_11_and_tildeP_12}),
\begin{eqnarray}
\tilde{\Pi}_{0}(\mathbf{k};q) & \equiv & k_{B}T \sum_{n} \, G_{0}(k+q) \, G_{0}(-k)
\nonumber \\
& = & \frac{1 - f_{F}(\xi(\mathbf{k})) - f_{F}(\xi(\mathbf{k}+\mathbf{q}))}{\xi(\mathbf{k}) + \xi(\mathbf{k}+\mathbf{q}) - i \Omega_{\nu}}
\label{tildeP_0}
\end{eqnarray}

\noindent
such that
\begin{equation}
\Pi_{0}(q) = \int \! \frac{d\mathbf{k}}{(2 \pi)^{3}} \, \tilde{\Pi}_{0}(\mathbf{k};q) \, ,
\label{integralP_0}
\end{equation}

\noindent
we obtain the following expression for the pair correlation function within the present approximation:
\begin{eqnarray}
g_{\uparrow \downarrow}(\boldsymbol{\rho}) & = & \int \! dq \, e^{i \Omega_{\nu} \eta} \, \Gamma_{0}(q)
\label{pair-correlation-function-above-Tc} \\
& \times & \int \! \frac{d\mathbf{k}}{(2 \pi)^{3}} \, e^{i \mathbf{k} \cdot \boldsymbol{\rho}} \, \tilde{\Pi}_{0}(\mathbf{k};q) \, 
                 \int \! \frac{d\mathbf{k}'}{(2 \pi)^{3}} \, e^{-i \mathbf{k}' \cdot \boldsymbol{\rho}} \, \tilde{\Pi}_{0}(\mathbf{k}';q) \, .
\nonumber 
\end{eqnarray}

\begin{figure}[h]
\begin{center}
\includegraphics[angle=0,width=7.5cm]{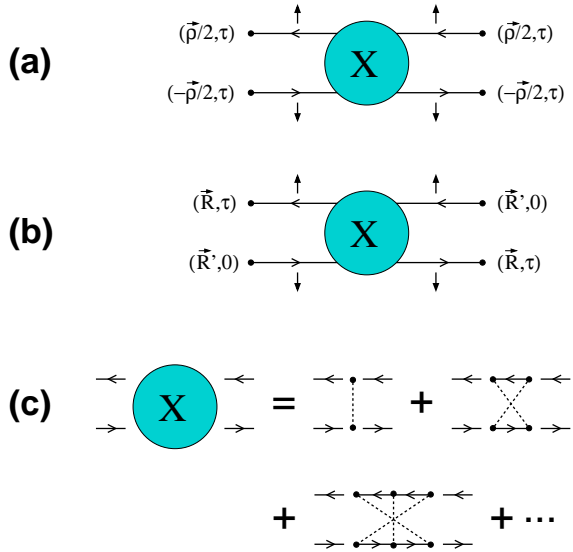}
\caption{(Color online) Schematic diagrammatic representation of (a) the pair correlation function (\ref{definition-pair-correlation-function}) and (b) the correlation function of the order parameter 
              (\ref{static-longitudinal-correlation-function}) (see below), where the space and imaginary time variables at the end points are indicated in each case. (c) Series of maximally crossed 
              diagrams $X$ which are used to approximate both correlation functions (\ref{definition-pair-correlation-function}) and (\ref{static-longitudinal-correlation-function}) above $T_{c}$.}
\label{fig-5}
\end{center}
\end{figure}

\noindent
The result (\ref{pair-correlation-function-above-Tc}) could have been obtained directly from the original expression (\ref{definition-pair-correlation-function}) in terms of the ordinary representation $\psi_{\sigma}(\mathbf{r})$ of the field operators which applies to the normal phase above $T_{c}$, provided one considers the series of ``maximally crossed diagrams'' depicted in Fig.÷\ref{fig-5}.

From the expression (\ref{pair-correlation-function-above-Tc}), we get for the volume integral of $g_{\uparrow \downarrow}(\boldsymbol{\rho})$
\begin{equation} 
\int \! d \boldsymbol{\rho} \, g_{\uparrow \downarrow}(\boldsymbol{\rho}) =
\int \! dq \, e^{i \Omega_{\nu} \eta} \,\Gamma_{0}(q) \! \int \! \frac{d\mathbf{k}}{(2 \pi)^{3}} \tilde{\Pi}_{0}(\mathbf{k};q)^{2} \, ,
\label{volume-integral-above-Tc}
\end{equation}

\noindent
and for its second moment
\begin{equation}
\int \! d \boldsymbol{\rho} \, \boldsymbol{\rho}^{2} \, g_{\uparrow \downarrow}(\boldsymbol{\rho}) = 
\!\! \int \! dq \, e^{i \Omega_{\nu} \eta} \, \Gamma_{0}(q) \!\!  \int \! \frac{d\mathbf{k}}{(2 \pi)^{3}} \! \left[ \nabla_{\mathbf{k}} \tilde{\Pi}_{0}(\mathbf{k};q) \right]^{2} \, ,
\label{second-moment-above-Tc}
\end{equation}

\noindent
from which $\xi_{\mathrm{pair}}$ can be obtained like in Eq.(\ref{xi-pair-definition}).

The leading short-range behavior of $g_{\uparrow \downarrow}(\boldsymbol{\rho})$ is somewhat simpler to obtain above than below $T_{c}$.
Similarly to the manipulations in Eq.(\ref{manipulation-1}), we now write in Eq.(\ref{pair-correlation-function-above-Tc}):
\begin{eqnarray}
& & \int \! \frac{d\mathbf{k}}{(2 \pi)^{3}} \, e^{i \mathbf{k} \cdot \boldsymbol{\rho}} \, \tilde{\Pi}_{0}(\mathbf{k};q) \,\,  _{\overrightarrow{(\boldsymbol{\rho}\rightarrow 0)}} \,\,
\int \! \frac{d\mathbf{k}}{(2 \pi)^{3}} \, e^{i \mathbf{k} \cdot \boldsymbol{\rho}} \, \frac{m}{\mathbf{k}^{2}} 
\nonumber \\
& + & \int \! \frac{d\mathbf{k}}{(2 \pi)^{3}} \, \left\{ \tilde{\Pi}_{0}(\mathbf{k};q) - \frac{m}{\mathbf{k}^{2}} \right\} \, = \, \frac{m}{4 \pi \rho} + R_{0}(q)
\label{manipulation-4}
\end{eqnarray}

\noindent
where
\begin{equation}
R_{0}(q) \equiv \Pi_{0}(q) -  \int \! \frac{d\mathbf{k}}{(2 \pi)^{3}} \, \frac{m}{\mathbf{k}^{2}} \, ,
\label{R_0-vs-Pi_0}
\end{equation}

\noindent
such that
\begin{eqnarray}
\Gamma_{0}(q) \, R_{0}(q) & = & - \frac{R_{0}(q)}{\frac{1}{v_{0}} + \Pi_{0}(q)} = - \frac{R_{0}(q)}{\frac{m}{4 \pi a_{F}} + R_{0}(q)}
\nonumber \\
& = & -1 - \frac{m}{4 \pi a_{F}} \, \Gamma_{0}(q)
\label{Gamma_0-vs-R_0}
\end{eqnarray}

\noindent
owing again to the regularization (\ref{ultraviolet-regularization}).
By entering the expression (\ref{manipulation-4}) in Eq.(\ref{pair-correlation-function-above-Tc}), we then obtain:
\begin{eqnarray}
& g_{\uparrow \downarrow}(\boldsymbol{\rho}) & \,\,  _{\overrightarrow{(\boldsymbol{\rho}\rightarrow 0)}} \,\,
\int \! dq \, e^{i \Omega_{\nu} \eta} \,\Gamma_{0}(q) \left[ \left(\frac{m}{4 \pi \rho}\right)^{2} + 2 \, \frac{m}{4 \pi \rho} \, R_{0}(q) \right]
\nonumber \\
& = & \left(\frac{m}{4 \pi}\right)^{2} \int \! dq \, e^{i \Omega_{\nu} \eta} \, \Gamma_{0}(q) \left( \frac{1}{\rho^{2}} - \frac{2}{a_{F} \rho} \right) 
\label{contact-fluctuations}
\end{eqnarray}

\noindent
where we now identify
\begin{equation}
\Delta_{\infty}^{2} \, = \int \! dq \, e^{i \Omega_{\nu} \eta} \, \Gamma_{0}(q) 
\label{Delta-infty-above}
\end{equation}

\noindent
such that $(m \, \Delta_{\infty})^{2}$ yields the Tan's contact $C$ above $T_{c}$ within the present theory \cite{PPS-2009}.

Correspondingly, the chemical potential is eliminated in favor of the density through the following expressions to which Eqs.(\ref{modified-density-equation})-(\ref{self-energy-below-Tc}) reduce above $T_{c}$:
\begin{equation}
n = 2 \int \! \frac{d\mathbf{k}}{(2 \pi)^{3}} \, k_{B}T \sum_{n} \, e^{i \omega_{n} \eta} \, G(\mathbf{k},\omega_{n})
\label{density-equation-above-Tc}
\end{equation}

\noindent
where
\begin{equation}
G(\mathbf{k},\omega_{n}) = \frac{1}{i \omega_{n} - \xi(\mathbf{k}) - \Sigma(\mathbf{k},\omega_{n})}
\label{modified-single-particle-Green-function}
\end{equation}

\noindent
with
\begin{equation}
\Sigma(k) = - \, \int \! dq \,\, \Gamma_{0}(q) \, G_{0}(q-k) \, .
\label{self-energy-above-Tc}
\end{equation}

\noindent
These expressions correspond to the non-self-consistent $t$-matrix approximation above $T_{c}$ in the form discussed in Ref.\cite{PPSC-2002} (see also Ref.\cite{NSR-1985}).

\begin{figure}[t]
\begin{center}
\includegraphics[angle=0,width=7.0cm]{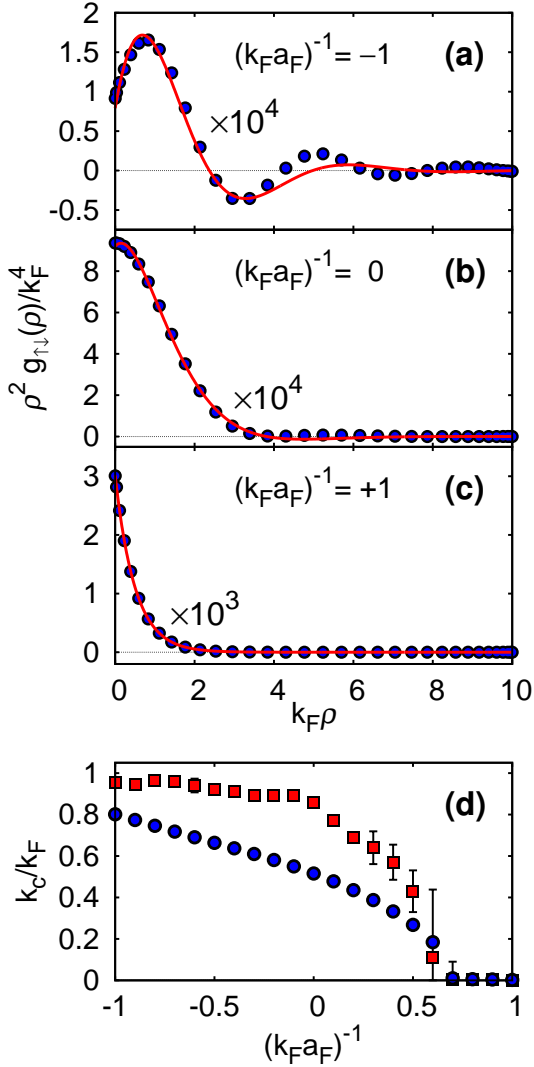}
\caption{(Color online) Radial profiles of the pair correlation function $g_{\uparrow \downarrow}(\boldsymbol{\rho})$ at $T_{c}$ multiplied by $\boldsymbol{\rho}^{2}$ for the couplings 
               $(k_{F} a_{F})^{-1}$: (a) $-1.0$, (b) $0.0$, (c) $+1.0$. Dots correspond to the expression (\ref{pair-correlation-function-above-Tc}) multiplied by $\boldsymbol{\rho}^{2}$ and lines represent
              the associated fits according to Eq.(\ref{fitting-function}).
              (d) The coupling dependence of the wave vector $k_{c}$ (dots), which characterizes the oscillating behavior of $g_{\uparrow \downarrow}(\boldsymbol{\rho})$ in Eq.(\ref{fitting-function}), 
              is compared at $T_{c}$ with that of the Luttinger wave vector $k_{L}$ (squares), which signals the presence of an underlying Fermi surface.}
\label{fig-6}
\end{center}
\end{figure}

Before presenting the numerical calculation of the expressions (\ref{volume-integral-above-Tc}) and (\ref{second-moment-above-Tc}) (and thus of $\xi_{\mathrm{pair}}$), 
it is worth considering in detail at least in some cases how the overall spatial dependence of $g_{\uparrow \downarrow}(\boldsymbol{\rho})$ evolves with coupling, from the BCS to the BEC 
regimes across unitarity.
This is shown in Fig.÷\ref{fig-6} for three couplings at the respective critical temperature.
In this figure, the results of the numerical calculation of the expression (\ref{pair-correlation-function-above-Tc}) multiplied by $\boldsymbol{\rho}^{2}$ (dots) are compared with the fits (lines) obtained
in terms of the following expression:
\begin{equation}
f(\rho) = A \cos(\phi_{0} + \sqrt{2} \rho k_{c}) \, e^{- \sqrt{2} \rho /\ell_{0}}
\label{fitting-function}
\end{equation} 

\noindent
where $\rho = |\boldsymbol{\rho}|$ and $(A,\phi_{0},k_{c},\ell_{0})$ are fitting parameters.
The numerical prefactors in the arguments of the cosine and of the exponential have been chosen in such a way that the wave vector $k_{c}$ which characterizes the oscillating behavior of
$g_{\uparrow \downarrow}(\boldsymbol{\rho})$ coincides with $k_{F}$ in the (extreme) BCS limit, while the length $\ell_{0}$ which characterizes the exponential decay of
$g_{\uparrow \downarrow}(\boldsymbol{\rho})$ coincides with $\xi_{\mathrm{pair}}=a_{F}/\sqrt{2}$ in the (extreme) BEC limit.
In addition, from these fits it turns out with good numerical accuracy that the product $A \cos(\phi_{0})$ coincides with $C/(16 \pi^{2})$ for all couplings, as expected from the result (\ref{contact-fluctuations}).

From Fig.÷\ref{fig-6} on the BCS side one notices a damped oscillating behavior with a period determined by the characteristic wave vector $k_{c}$ of Eq.(\ref{fitting-function}).
On physical grounds, one expects $k_{c}$ to be related to the radius $k_{L}$ of the underlying Fermi surface, which can, in turn, be identified from the dispersion relations associated with the single-particle spectral function \cite{Camerino-Jila-2011}. 
As a consequence, the oscillating behavior of the pair correlation function is bound to disappear on the BEC side of unitarity once the underlying Fermi surface has collapsed, a situation which corresponds to panel (c) of Fig.÷\ref{fig-6}.
The dependence of the wave vector $k_{c}$ on coupling obtained in this way at the respective critical temperatures is shown in panel (d) of Fig.÷\ref{fig-6}, where it is also compared with the corresponding dependence of the Luttinger wave vector $k_{L}$ as reported in Ref.\cite{Camerino-Jila-2011}.
Note from this plot that, as expected, $k_{c}$ and $k_{L}$ both vanish at the same coupling value ($\simeq 0.6$).
We have also verified that, in all cases, the function $g_{\uparrow \downarrow}(\boldsymbol{\rho}) + (n/2)^{2}$ remains positive.
This represents a non-trivial test on our approximate theory since this function, being by definition a probability distribution, has to remain non-negative for all $\boldsymbol{\rho}$.

\begin{figure}[h]
\begin{center}
\includegraphics[angle=0,width=6.8cm]{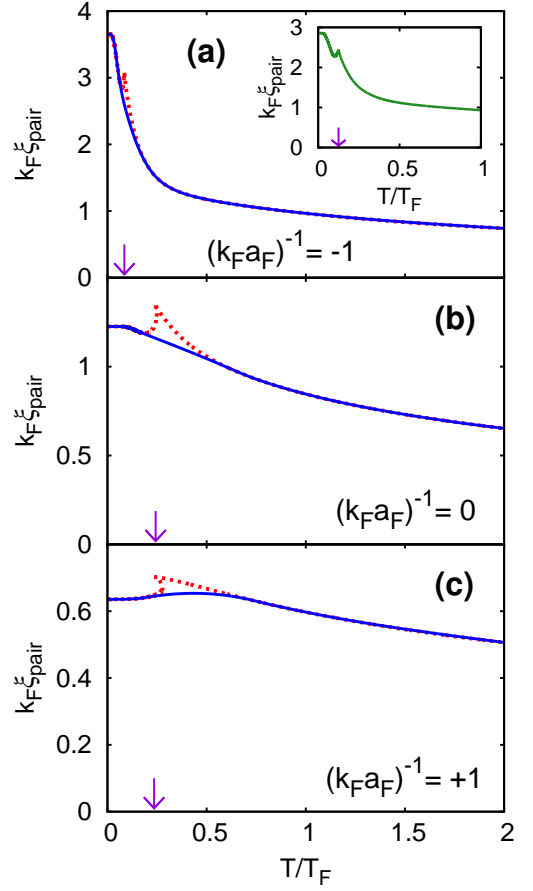}
\caption{(Color online) The temperature dependence of $\xi_{\mathrm{pair}}$, obtained with the inclusion of pairing fluctuations both below and above $T_{c}$, is shown for the couplings 
              $(k_{F} a_{F})^{-1}$: (a) $-1.0$, (b) $0.0$, (c) $+1.0$. The arrows locate the corresponding critical temperatures. Dotted lines correspond to the numerical results of the calculation, 
              while full lines represent an interpolation which smooths out the cusp-like feature present in the numerical results close to $T_{c}$. The inset of panel (a) shows the temperature dependence of
              $\xi_{\mathrm{pair}}$ for $(k_{F} a_{F})^{-1} = -1.0$, which is obtained from the expressions (\ref{pair-correlation-function-below-Tc}) below $T_{c}$ and (\ref{pair-correlation-function-above-Tc}) 
              above $T_{c}$, but now with $\Delta$ and $\mu$ at the mean-field level.}
\label{fig-7}
\end{center}
\end{figure}

The complete temperature dependence of $\xi_{\mathrm{pair}}$, both below and above $T_{c}$, is reported in Fig.÷\ref{fig-7} for the same couplings of Fig.÷\ref{fig-6}.
Here, the bare results of the calculation (dotted lines) have been further interpolated (full lines) so as to smooth out the cusp-like feature which is present in all cases close to $T_{c}$ (the difference
between the original and the smoothed data never exceeding $10 \%$ in practice).
With this smoothing provision, the overall behavior of $\xi_{\mathrm{pair}}$ corresponds basically to a decreasing function of temperature.
Note also that, in contrast to the mean-field case reported in Fig.÷\ref{fig-2}(b), once pairing fluctuations are included no universal behavior is obtained from the smoothed curves of Fig.÷\ref{fig-7} at different couplings by a suitable rescaling of the variables.

One might be tempted to conclude that the cusp-like feature present in the unsmoothed data of Fig.÷\ref{fig-7} should be attributed to the occurrence of a reentrant behavior of the gap parameter in the vicinity of $T_{c}$.
This feature occurs in the $t$-matrix approaches, both in their non-self-consistent \cite{PPS-2004} and self-consistent \cite{Zwerger-2007} versions, where it is known to affect several thermodynamic quantities in a similar fashion to that shown in Fig.÷\ref{fig-7}.
However, we have explicitly verified that this feature shows up in the temperature dependence of $\xi_{\mathrm{pair}}$ also when the gap parameter is taken at the mean-field level for which no reentrant behavior occurs.
This is shown in the inset of Fig.÷\ref{fig-7}(a) for the BCS side of unitarity.
A similar abrupt behavior when crossing $T_{c}$ is known to occur at the BCS mean-field level for other thermodynamic quantities as well \cite{Tinkham-1980}.

\begin{figure}[t]
\begin{center}
\includegraphics[angle=0,width=7.0cm]{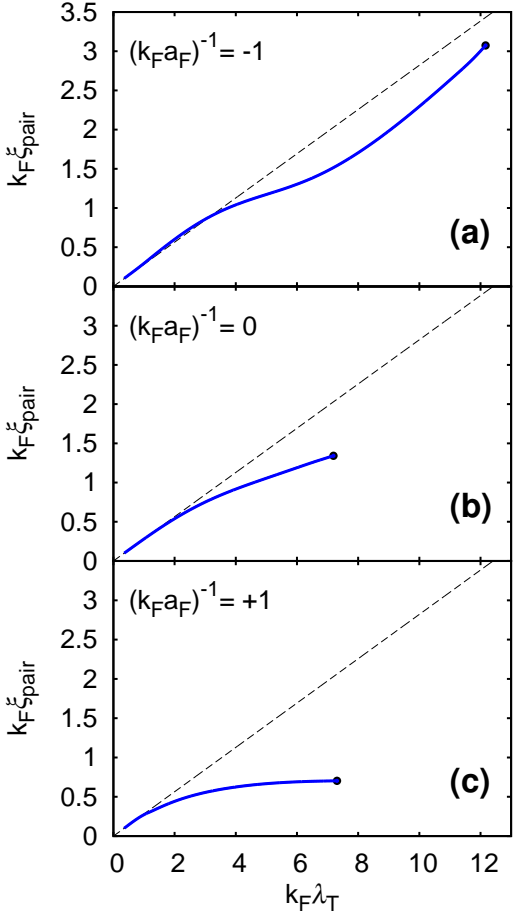}
\caption{(Color online) The temperature dependence of $\xi_{\mathrm{pair}}$ with the inclusion of pairing fluctuations above $T_{c}$ (full lines) is expressed in terms of the thermal wavelength 
              $\lambda_{\mathrm{T}}$, for the couplings $(k_{F} a_{F})^{-1}$: (a) $-1.0$, (b) $0.0$, (c) $+1.0$. The dashed lines correspond to the asymptotic value $0.28 \, \lambda_{\mathrm{T}}$, 
              while the dots mark the onset of the superfluid phase in each case.}
\label{fig-8}
\end{center}
\end{figure}

An interesting feature which results from the above temperature dependence of $\xi_{\mathrm{pair}}$ is that, above but close to $T_{c}$, this dependence is \emph{steeper} on the BCS than on the BEC side of the crossover (while in all cases when $T \gg T_{c}$ it decays rather slowly like $1/\sqrt{T}$ - see below).
This feature will be exploited in subsection II-E when comparing with available experimental data related to $\xi_{\mathrm{pair}}$ in the normal phase.
From the above results it also appears that for an attractive Fermi gas intra-pair correlations begin to build up in a substantial way already at temperatures of the order of the Fermi temperature $T_{F}$.
Correspondingly, inter-pair correlations, which establish the (off-diagonal) long-range order below $T_{c}$, will be seen in the next Section to become effective above $T_{c}$ only at lower temperatures.

Finally, Fig.÷\ref{fig-8} recasts the temperature dependence of $\xi_{\mathrm{pair}}$ above $T_{c}$ in terms of the thermal wavelength $\lambda_{\mathrm{T}} = \sqrt{ \frac{2 \pi}{m k_{B} T} }$ 
(defined like in Ref.\cite{Huang-1963}).
In each panel, the straight (dashed) line correspond to the asymptotic value $\lambda_{\mathrm{T}}/\sqrt{4 \pi} \simeq 0.28 \, \lambda_{\mathrm{T}}$ which is reached by $\xi_{\mathrm{pair}}$ in the high-temperature (classical) limit irrespective of coupling, a result that can be obtained analytically from the expressions (\ref{volume-integral-above-Tc}) and (\ref{second-moment-above-Tc}) (cf. Appendix B).
Once this asymptotic value is reached, pair correlations can be considered to have been completely overcome by thermal fluctuations.
Also in the context of Fig.÷\ref{fig-8}, the dependence of $\xi_{\mathrm{pair}}$ on temperature just above $T_{c}$ appears more marked on the BCS with respect to the BEC side of unitarity.

\vspace{0.1cm}
\begin{center}
{\bf E. Comparison with available experimental data on the proximity effect in the normal phase}
\end{center}
\vspace{0.1cm}

The results of subsection II-D, about the temperature dependence of $\xi_{\mathrm{pair}}$ in the normal phase above $T_{c}$ for various couplings, can be related to recent measurements about the temperature dependence of the normal coherence length $\xi_{\mathrm{N}}$.
This dependence was obtained from the proximity effect in an SS'S superconducting Josephson junction made of high-temperature (cuprate) superconducting materials, with the barrier region S' constrained to the normal phase \cite{KK-2013}.

\begin{figure}[h]
\begin{center}
\includegraphics[angle=0,width=7.5cm]{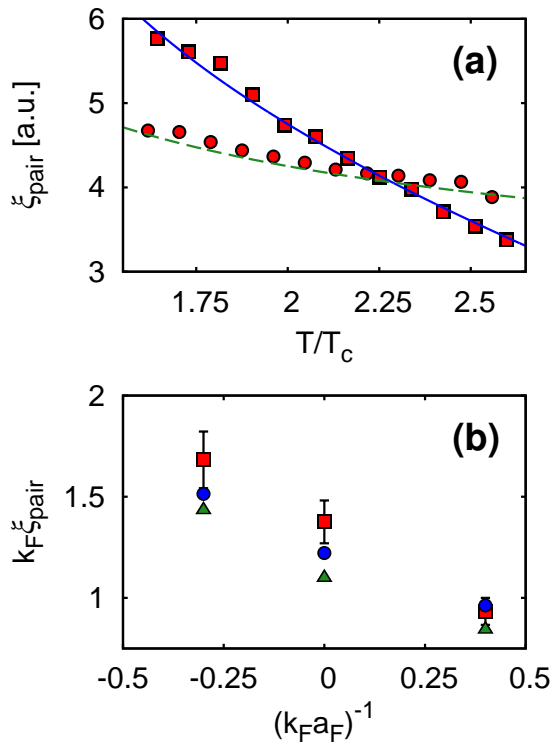}
\caption{(Color online) (a) Comparison between the temperature dependence of $\xi_{\mathrm{pair}}$ above $T_{c}$ (lines) and of $\xi_{\mathrm{N}}$ obtained experimentally in Ref.\cite{KK-2013}, 
               for an optimally-doped (LSCO-0.18, squares) and for an under-doped (LSCO-0.10, circles) material. 
               (b) The experimental data for $\xi_{\mathrm{pair}}$ obtained in Ref.\cite{Ketterle-2008} from radio-frequency spectroscopy of an ultra-cold Fermi gas taken at the temperatures $T=(0.1,0.1,0.2)T_{F}$ 
               from left to right (squares with error bars) are compared with our calculated values, both at the $T=0$ mean-field level (triangles) and with the inclusion of pairing fluctuations at the same 
               temperatures of the experiment (circles).}
\label{fig-9}
\end{center}
\end{figure}

Specifically, it was found in Ref.\cite{KK-2013} that the temperature dependence of $\xi_{\mathrm{N}}$ is somewhat steeper for an optimally-doped with respect to an under-doped material, a result that was attributed to the presence of pre-formed pairs in the pseudo-gap regime of the cuprate barrier. 
This result appears to be in line with our finding that the temperature dependence of $\xi_{\mathrm{pair}}$ is steeper on the BCS than on the BEC side of the crossover, \emph{provided} one associates $\xi_{\mathrm{N}}$ with $\xi_{\mathrm{pair}}$ and attributes a stronger coupling to the under-doped with respect to the optimally-doped regime of the high-temperature (cuprate) superconductors utilized in the experiment.
(A more extensive discussion about the relationship between $\xi_{\mathrm{N}}$ and $\xi_{\mathrm{pair}}$ is reported in Appendix C.)

It should be further pointed out in this context that the data of Ref.\cite{KK-2013} on the proximity effect refer specifically to superconducting properties above $T_{c}$, which are not bound to survive up to the crossover temperature at which pseudo-gap phenomena eventually disappear (as clearly reported in Fig.5 of Ref.\cite{KK-2013}).
For this reason, the findings of Ref.\cite{KK-2013}, together with our interpretation here in terms of pairing fluctuations above $T_{c}$, are not necessarily in contrast with theories involving competing order parameters of a different kind.

In Fig.÷\ref{fig-9}(a) we quantify the comparison, between the experimental temperature dependence of $\xi_{\mathrm{N}}$ reported in Ref.\cite{KK-2013} and our theoretical
temperature dependence of $\xi_{\mathrm{pair}}$ in the normal phase.
This is done by (i) rescaling both $\xi_{\mathrm{N}}$ and $\xi_{\mathrm{pair}}$ to arbitrary units so that they coincide with each other 
at about $T = 2 T_{c}$ in both cases analyzed experimentally, and (ii) varying the coupling $(k_{F} a_{F})^{-1}$ at which the temperature dependence of $\xi_{\mathrm{pair}}$ is calculated 
until agreement is found with the corresponding temperature dependence of $\xi_{\mathrm{N}}$.
As shown in Fig.÷\ref{fig-9}(a), we find in this way a reasonably good agreement between the experimental and theoretical temperature dependence of these lengths, provided we attribute to the 
under-doped material a coupling value of about $-0.4$ close to unitarity, and to the optimally-doped material a coupling value of about $-3.5$ well inside the BCS regime.
(We have verified that the latter value shifts to $-3.0$ when the Gor'kov-Melik-Barkhudarov correction is further included \cite{G-MB-1961}.)
In absolute units of $k_{F}^{-1}$, to the above coupling values $-0.4$ and $-3.5$ there correspond the values $k_{F} \xi_{\mathrm{pair}} \simeq (4.3,40)$, in order, at the lowest temperature of about $1.5 T_{c}$ at which the measurements were taken.

An additional comparison with the experimental data involving $\xi_{\mathrm{pair}}$, which can be done \emph{directly} in absolute units of $k_{F}^{-1}$, is reported in Fig.÷\ref{fig-9}(b).
Here, the experimental data for $\xi_{\mathrm{pair}}$ obtained from radio-frequency spectroscopy of an ultra-cold Fermi gas at finite temperatures \cite{Ketterle-2008} (squares) are compared with our calculations.
This comparison shows that the combined effect of temperature and pairing fluctuations over and above mean-field (circles) results in a closer agreement with the experimental data with respect to the mean-field results taken at $T=0$ (triangles).  
[We remark that, to obtain this comparison, the bare experimental data obtained in Ref.\cite{Ketterle-2008} from the widths of the radio-frequency spectra have been suitably converted into values of 
$\xi_{\mathrm{pair}}$, utilizing a prescription given in the inset of Fig.1(c) of Ref.\cite{Ketterle-2008} itself.]

\vspace{-0.2cm}
\section{III. The correlation function of the order parameter and the associated length $\xi_{\mathrm{phase}}$} 
\label{sec: xi-phase}

In this Section, we examine the intra-pair correlations which become critical when approaching $T_{c}$ from above and are thus responsible for the building up of the superconducting (off-diagonal) long-range order.

To this end, we will retrace the treatment of Ref.\cite{PS-1996}, where the (longitudinal) correlation function of the order parameter was determined below $T_{c}$ from a functional-integral approach, and rephrase it in terms of a diagrammatic approach from which the inter-pair (healing) length $\xi_{\mathrm{phase}}$ will be obtained \emph{both below and above} $T_{c}$ throughout the BCS-BEC crossover (while in Ref.\cite{PS-1996} $\xi_{\mathrm{phase}}$ was calculated at $T=0$ only). 
Besides being somewhat simpler to handle than the functional-integral approach, the diagrammatic approach used here for the correlation function of the order parameter has the advantage of being formally related to that describing the pair correlation function utilized in the previous Section.

\vspace{0.1cm}
\begin{center}
{\bf A. Results below $T_{c}$ for the ``longitudinal'' component of the correlation function}
\end{center}
\vspace{0.1cm}

We begin by considering the superfluid phase, where (with reference to the direction of broken symmetry) one needs to distinguish between the longitudinal and transverse components of the correlation function of the order parameter. 
Since only to the longitudinal component one can associate a finite value of the healing length for inter-pair correlation, in the following we shall deal with this component only.

In terms of the center-of-mass coordinates of the pairs, we thus define a ``longitudinal pair operator''
\begin{equation}
\varphi_{\parallel}(\mathbf{R}) = \frac{1}{2 |\Delta|} \left[ \Delta^{*} \varphi(\mathbf{R}) + \Delta \varphi^{\dagger}(\mathbf{R}) \right]
\label{longitudinal-operator}
\end{equation}

\noindent
where $\varphi(\mathbf{R}) = v_{0} \psi_{\downarrow}(\mathbf{R}) \psi_{\uparrow}(\mathbf{R})$, such that $\langle \varphi(\mathbf{R}) \rangle = \Delta$ for the homogeneous system we are considering.
Here, $v_{0}$ is the strength of the inter-particle attractive potential, which will be taken to vanish according to the regularization (\ref{ultraviolet-regularization}) \emph{only} at the end of the calculation.  
(We shall also take eventually $\Delta$ to be real without loss of generality.)

Following Ref.\cite{PS-1996}, we consider the \emph{static longitudinal correlation function} of the order parameter defined by:
\begin{equation}
F_{\parallel}(\mathbf{R} - \mathbf{R}') = \int_{0}^{\beta} \! d\tau \, \langle T_{\tau} \left[ \varphi_{\parallel}(\mathbf{R,\tau}) \varphi_{\parallel}(\mathbf{R}',\tau=0) \right] \rangle - \beta \, |\Delta|^{2}
\label{static-longitudinal-correlation-function}
\end{equation}

\noindent
where $\beta = (k_{B} T)^{-1}$ is the inverse temperature.
With reference to the Nambu representation of the field operators and to the general expression (\ref{Bethe-Salpeter-equation}) of the two-particle Green's function, the correlation function (\ref{static-longitudinal-correlation-function}) can be rewritten as follows:
\begin{eqnarray}
& & F_{\parallel}(\mathbf{R} - \mathbf{R}') = \frac{v_{0}^{2}}{4 |\Delta|^{2}}  \int_{0}^{\beta} \! d\tau \, \left\{ (\Delta^{*})^{2} \, \mathcal{G}_{2}(1,2;1'^{+},2'^{+}) \right.
\nonumber \\
& + & \Delta^{*} \Delta \, \mathcal{G}_{2}(1,2';1'^{+},2^{+}) + \Delta \Delta^{*} \, \mathcal{G}_{2}(1',2;1^{+},2'^{+})
\nonumber \\
& + & \left. \Delta^{2} \, \mathcal{G}_{2}(1',2';1^{+},2^{+}) \right\} - \beta \, |\Delta|^{2}
\label{static-longitudinal-correlation-function-Nambu}
\end{eqnarray}

\noindent 
where now
\begin{eqnarray}
1 & = & (\mathbf{R}, \tau, \ell=1)           \nonumber \\
2 & = & (\mathbf{R}', \tau=0, \ell=1)      \nonumber \\
1' & = & (\mathbf{R}, \tau, \ell=2)          \nonumber \\
2' & = & (\mathbf{R}', \tau=0, \ell=2) 
\label{compact-notation-2}
\end{eqnarray}

\noindent
is the relevant ``dictionary'' to be applied to the correlation function of the order parameter.

Akin to the treatment of the pair correlation function that was made in Section II, only the first two terms on the right-hand side of Eq.(\ref{Bethe-Salpeter-equation}) contribute within the BCS (mean-field) decoupling, yielding:
\begin{eqnarray}
F_{\parallel}(\mathbf{R} - \mathbf{R}') & = &  - \frac{v_{0}}{2} \, \delta(\mathbf{R} - \mathbf{R}') - \frac{v_{0}^{2}}{2} \, \int \! \frac{d\mathbf{Q}}{(2 \pi)^{3}} \, 
e^{i \mathbf{Q} \cdot (\mathbf{R}-\mathbf{R}')} 
\nonumber \\
& \times & \left\{ A(\mathbf{Q},\Omega_{\nu}=0) + B(\mathbf{Q},\Omega_{\nu}=0) \right\}
\label{F-parallel-within-mean-field}
\end{eqnarray}

\noindent
where $A(q)$ and $B(q)$ are the same quantities of Eq.(\ref{T-matrix-elements}).
The expression (\ref{F-parallel-within-mean-field}), however, does not survive the regularization (\ref{ultraviolet-regularization}) when $v_{0} \rightarrow 0$ and will therefore be neglected in the following.

Quite generally, the remaining term for $\mathcal{G}_{2}$ in Eq.(\ref{Bethe-Salpeter-equation}), which contains the many-particle T-matrix of Fig.÷\ref{fig-3}(a), gives the following contributions to the correlation function (\ref{static-longitudinal-correlation-function-Nambu}):
\begin{eqnarray}
& & F_{\parallel}(\mathbf{R} - \mathbf{R}') = - \frac{v_{0}^{2}}{4} \! \int \! \frac{d\mathbf{Q}}{(2 \pi)^{3}} e^{i \mathbf{Q} \cdot (\mathbf{R}-\mathbf{R}')} \!\!\!\!
\sum_{\ell_{3} \ell_{4} \ell_{5} \ell_{6}} \! T^{\ell_{3} \ell_{4}}_{\ell_{6} \ell_{5}}(q)
\nonumber \\
& \times & \left\{ \int \!\! dk \, \mathcal{G}_{1 \ell_{3}}(k+q) \mathcal{G}_{\ell_{6} 2}(k) \!\! \int \!\! dk' \, \mathcal{G}_{\ell_{4} 2}(k'+q) \mathcal{G}_{1 \ell_{5}}(k') \right.
\nonumber \\
& + &  \int \!\! dk \,\mathcal{G}_{1 \ell_{3}}(k+q) \mathcal{G}_{\ell_{6} 2}(k) \!\! \int \!\! dk' \,\mathcal{G}_{\ell_{4} 1}(k'+q) \mathcal{G}_{2 \ell_{5}}(k')
\nonumber \\
& + &  \int \!\! dk \, \mathcal{G}_{2 \ell_{3}}(k+q) \mathcal{G}_{\ell_{6} 1}(k) \!\! \int \!\! dk' \, \mathcal{G}_{\ell_{4} 2}(k'+q) \mathcal{G}_{1 \ell_{5}}(k')
\nonumber \\
& + & \left. \!\!\! \int \!\! dk \, \mathcal{G}_{2 \ell_{3}}(k+q) \mathcal{G}_{\ell_{6} 1}(k) \!\! \int \!\! dk' \, \mathcal{G}_{\ell_{4} 1}(k'+q) \mathcal{G}_{2 \ell_{5}}(k') \!\! \right\}
\label{F-parallel-beyond-mean-field}
\end{eqnarray}

\noindent
where $q=(\mathbf{Q},\Omega_{\nu}=0)$ has to be understood in the above expression whenever it appears.

As we did in subsection II-C, we again limit ourselves to consider the series of ladder diagrams in the broken-symmetry phase for an inter-particle interaction of the contact type, which are depicted 
in Fig.÷\ref{fig-3}(a).
After a long but straightforward calculation, in the relevant limit when $v_{0} \rightarrow 0$ the expression (\ref{F-parallel-beyond-mean-field}) reduces eventually to the result:
\begin{equation}
F_{\parallel}(\mathbf{R} - \mathbf{R}') = \frac{1}{2} \! \int \! \frac{d\mathbf{Q}}{(2 \pi)^{3}} \,
\frac{ e^{i \mathbf{Q} \cdot (\mathbf{R}-\mathbf{R}')}}{A(\mathbf{Q},\Omega_{\nu}=0) + B(\mathbf{Q},\Omega_{\nu}=0)} 
\label{F-parallel-beyond-mean-field-final}
\end{equation}

\noindent
which coincides with that obtained originally in Ref.\cite{PS-1996} through a functional-integral approach at the one-loop order (Gaussian fluctuations).

The novelty here is that the diagrammatic structures of the correlation function of the order parameter (\ref{static-longitudinal-correlation-function}) and of the pair correlation function
(\ref{definition-pair-correlation-function}) have been treated on equal footing (a feature which appears mostly evident when dealing with the normal phase above $T_{c}$, as it was already remarked when
drawing the diagrams of Fig.÷\ref{fig-5}).
For this reason, the values of $\Delta$ and $\mu$ to be used in the numerical calculation of the expression (\ref{F-parallel-beyond-mean-field-final}) should be taken in line with the treatment of subsection II-C which includes pairing fluctuations below $T_{c}$, although in Ref.\cite{PS-1996} they where considered at the mean-field level like in subsection II-B (and also at zero temperature only).

Nevertheless, it is of interest to calculate the coupling and temperature dependence of the healing length $\xi_{\mathrm{phase}}$ extracted from Eq.(\ref{F-parallel-beyond-mean-field-final}) 
below $T_{c}$ also with the values of $\Delta$ and $\mu$ taken at the mean-field level, since this allows us to compare with the results of Ref.\cite{SPS-2013} where the same quantity was extracted from the spatial profiles of the order parameter obtained by a numerical solution of the Bogoliubov-de Gennes (BdG) equations for an isolated vortex embedded in an infinite superfluid. 

\begin{figure}[h]
\begin{center}
\includegraphics[angle=0,width=7.0cm]{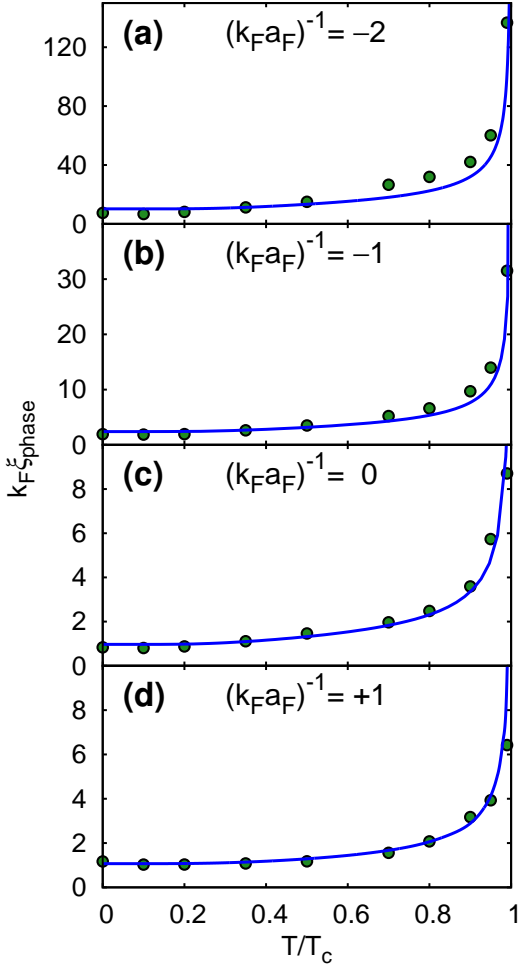}
\caption{(Color online) The temperature dependence of $\xi_{\mathrm{phase}}$ obtained from Eq.(\ref{small-Q-expansion}) with $\Delta$ and $\mu$ taken at the mean-field level (full lines), 
              is compared with the results reported in Fig.9 of Ref.\cite{SPS-2013} (dots) which were obtained by a numerical solution of the BdG equations, for the couplings $(k_{F} a_{F})^{-1}$: 
              (a) $-2.0$, (b) $-1.0$, (c) $0.0$, (d) $+1.0$.
              Here, the results of the present calculation have been rescaled by an overall factor of $2/3$, which takes into account the different definitions used for the same 
              physical quantity by the two independent numerical calculations.}
\label{fig-10}
\end{center}
\end{figure}

Quite generally, to obtain the value of the healing length $\xi_{\mathrm{phase}}$ we follow the procedure adopted in Ref.\cite{PS-1996} and expand the quantity $A(\mathbf{Q},\Omega_{\nu}=0) + B(\mathbf{Q},\Omega_{\nu}=0)$ in the integrand of Eq.(\ref{F-parallel-beyond-mean-field-final}) for small values of $\mathbf{Q}$:
\begin{equation}
A(\mathbf{Q},\Omega_{\nu}=0) + B(\mathbf{Q},\Omega_{\nu}=0) = a + b \, \mathbf{Q}^{2} + \cdots
\label{small-Q-expansion}
\end{equation}

\noindent
in such a way that $\xi_{\mathrm{phase}} = \sqrt{b / a}$ \cite{footnote-1}.
In addition, to account for the independent definitions used for $\xi_{\mathrm{pair}}$ and $\xi_{\mathrm{phase}}$, in the following we adopt the convention of rescaling the value of
$\xi_{\mathrm{phase}}$ obtained from the expression (\ref{small-Q-expansion}) at $T=0$ in the BCS limit $(k_{F} a_{F})^{-1} \ll -1$ in such a way that it coincides with the value of $\xi_{\mathrm{pair}}$ in that limit (as one would expect it to be the case on physical grounds). 
In this way, we set $\xi_{\mathrm{phase}} = \frac{3}{\sqrt{2}} \sqrt{\frac{b}{a}}$ \cite{PPS-2010}.

At the mean-field level, the condition $A(\mathbf{Q}=0,\Omega_{\nu}=0) = B(\mathbf{Q}=0,\Omega_{\nu}=0)$ is equivalent to the BCS gap equation (\ref{BCS-gap-equation}).
Upon approaching the critical temperature from below, $B(\mathbf{Q}=0,\Omega_{\nu}=0)$ given by Eq.(\ref{P_11_and_P_12}) vanishes like 
$\Delta^{2}(T \rightarrow T_{c}^{-}) \propto (T_{c} - T)$.
This implies that also the coefficient $a$ of Eq.(\ref{small-Q-expansion}) vanishes like $(T_{c} - T)$ in this limit, such that $\xi_{\mathrm{phase}} \propto (T_{c} - T)^{-1/2}$ consistent with the value $1/2$ of the mean-field critical exponent.

Figure \ref{fig-10} compares over the temperature interval from $T=0$ to $T_{c}$ the results of the present calculation for $\xi_{\mathrm{phase}}$ (whereby pairing fluctuations beyond the BCS decoupling are included in the broken-symmetry phase with a \emph{homogeneous} gap parameter $\Delta$ through the series of ladder diagrams depicted in Fig.÷\ref{fig-3}(a), where the values of $\Delta$ and $\mu$ are taken at the mean-field level), with the results obtained alternatively in Ref.\cite{SPS-2013} by a numerical solution of the BdG equations with a \emph{spatially dependent} $\Delta$ that represents an isolated vortex embedded in an infinite superfluid.
The rather remarkable agreement between these two independent calculations confirms one's expectation that a mean-field calculation for an inhomogeneous situation (of the type usually dealt with by the BdG equations) can actually contain contributions from what would usually be referred to as fluctuation corrections in a homogeneous situation.
This is in line with a general consideration that, in an inhomogeneous situation, the imprint of the quasiparticle spectrum can be found in the ground-state wave function \cite{Gross-1963}.

\begin{figure}[h]
\begin{center}
\includegraphics[angle=0,width=8.0cm]{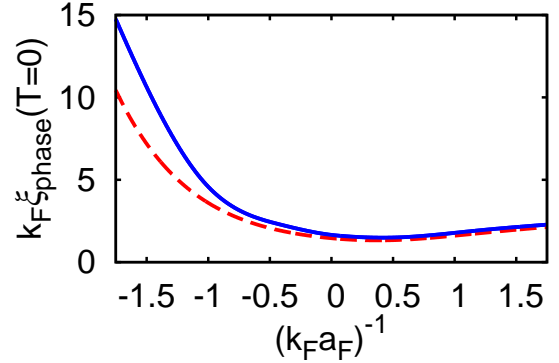}
\caption{(Color online) Coupling dependence of $\xi_{\mathrm{phase}}$ at $T=0$, when the values of $\Delta$ and $\mu$ on which it depends include (full line) or do not include (dashed line) 
              pairing fluctuations.}
\label{fig-11}
\end{center}
\end{figure}

The numerical values of $\xi_{\mathrm{phase}}$ somewhat change when the values of $\Delta$ and $\mu$ to be inserted into the correlation function (\ref{F-parallel-beyond-mean-field-final}) are instead obtained by including also pairing fluctuations (cf. subsection II-C).
A comparison between the coupling dependence of $\xi_{\mathrm{phase}}$ at $T=0$, obtained when the values of $\Delta$ and $\mu$ include or do not include pairing fluctuations, is shown in
Fig.÷\ref{fig-11}. 
Note that the use of values of $\Delta$ and $\mu$ beyond mean field somewhat  increases $\xi_{\mathrm{phase}}$ on the BCS side of unitarity.
This is in line with the fact that the inclusion of pairing fluctuations, to the extent that it decreases the value of the critical temperature at a given coupling (see Fig.÷\ref{fig-13} below), has the effect of re-normalizing the coupling to a smaller value along similar lines to what was already pointed out in the discussion of Fig.4(a).

\vspace{0.1cm}
\begin{center}
{\bf B. Results above $T_{c}$ and the crossover temperature $T^{*}$}
\end{center}
\vspace{0.1cm}

In the normal phase, $B(q)=0$ and $A(q)=-1/v_{0}-\Pi_{0}(q)= 1 / \Gamma_{0}(q)$ with the notation of subsection II-D.
Correspondingly, the expression (\ref{F-parallel-beyond-mean-field-final}) reduces to:
\begin{equation}
F(\mathbf{R} - \mathbf{R}') = \frac{1}{2} \! \int \! \frac{d\mathbf{Q}}{(2 \pi)^{3}} \,
e^{i \mathbf{Q} \cdot (\mathbf{R}-\mathbf{R}')} \, \Gamma_{0}(\mathbf{Q},\Omega_{\nu}=0) 
\label{F-above-Tc}
\end{equation}

\noindent
where the suffix $_{\parallel}$ has been dropped since in the normal phase no reference remains to the direction of broken symmetry.
As it was mentioned in the previous subsection, above $T_{c}$ it is possible to appreciate most readily that the difference between the pair correlation function [Eq.(\ref{pair-correlation-function-above-Tc})] and the correlation function of the order parameter [Eq.(\ref{F-above-Tc})] is due to the ways the external variables are set in the diagrammatic structure of Fig.÷\ref{fig-5}, which select alternatively the intra-pair variable $\boldsymbol{\rho}$ or the inter-pair variable $\mathbf{R} - \mathbf{R}'$.

\begin{figure}[h]
\begin{center}
\includegraphics[angle=0,width=7.0cm]{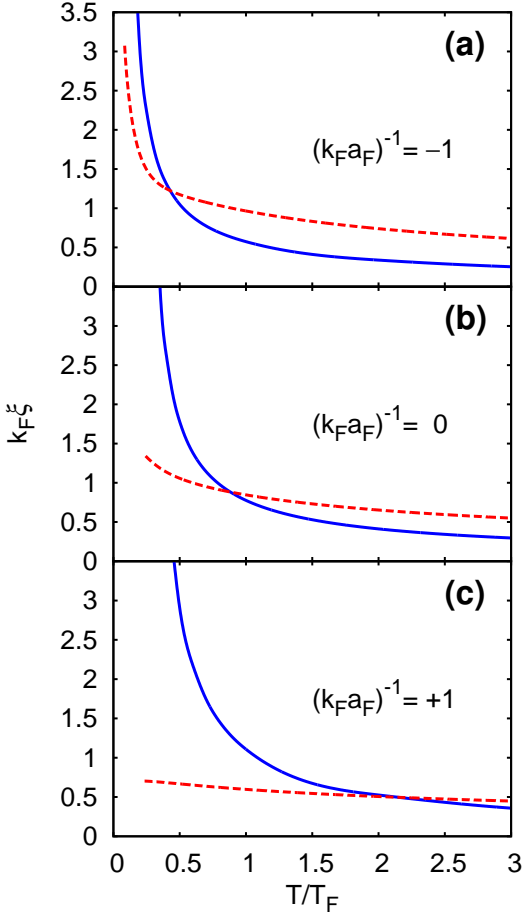}
\caption{(Color online) The temperature dependence of $\xi_{\mathrm{phase}}$ (full lines) is compared with that of $\xi_{\mathrm{pair}}$ (dashed lines) above $T_{c}$ for the couplings $(k_{F} a_{F})^{-1}$: 
             (a) $-1.0$, (b) $0.0$, (c) $+1.0$.}
\label{fig-12}
\end{center}
\end{figure}

The expression (\ref{F-above-Tc}) holds at any temperature above $T_{c}$, and $\xi_{\mathrm{phase}}$ can correspondingly be obtained by an expansion similar to Eq.(\ref{small-Q-expansion}).
The temperature dependence of $\xi_{\mathrm{phase}}$ obtained in this way for three characteristic couplings across the BCS-BEC crossover is shown in Fig.÷\ref{fig-12}, where it is also compared with that of $\xi_{\mathrm{pair}}$ above $T_{c}$ reported previously in Fig.÷\ref{fig-7}.

From these plots one notices a steeper temperature dependence of $\xi_{\mathrm{phase}}$ with respect to $\xi_{\mathrm{pair}}$, which at any coupling leads to a \emph{crossing} of the corresponding curves at a characteristic temperature $T^{*}$.
This temperature, which can be thus obtained for all couplings throughout the BCS-BEC crossover, has then the meaning of a ``crossover temperature'' below which inter-pair correlations begin to be built from the intra-pair correlations that are already present above this temperature.
[In Appendix B, we shall verify that, in the high-temperature limit $T \gtrsim T_{F}$, $\xi_{\mathrm{phase}}$ decays like $[(T/T_{F}) \ln (T/T_{F})]^{-1/2}$ and therefore at a faster rate than
$\xi_{\mathrm{pair}}$ which instead decays like $[(T/T_{F})]^{-1/2}$.] 

\begin{figure}[h]
\begin{center}
\includegraphics[angle=0,width=8.5cm]{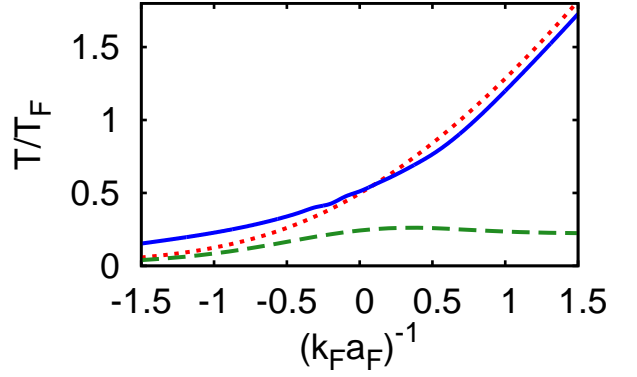}
\caption{(Color online) The coupling dependence of $T^{*}$, which results from the choice $\xi_{\mathrm{phase}}(T^{*})/\xi_{\mathrm{pair}}(T^{*})=\pi^{2}/6$ (full line), is compared with that of the
              mean-field critical temperature $T_{c}^{\mathrm{BCS}}$ (dotted line). The critical temperature $T_{c}$ within the $t$-matrix approximation is also reported for comparison (dashed line).}
\label{fig-13}
\end{center}
\end{figure}

Being a crossover temperature, the precise value of $T^{*}$ at a given coupling does not matter, on physical grounds the only reasonable condition being that the ratio 
$\xi_{\mathrm{phase}}(T^{*})/\xi_{\mathrm{pair}}(T^{*})$ remains of order unity at $T^{*}$.
Interestingly enough, we have found that with the choice $\xi_{\mathrm{phase}}(T^{*})/\xi_{\mathrm{pair}}(T^{*})=\pi^{2}/6\simeq1.64$ the overall coupling dependence of $T^{*}$ results quite similar to that of the BCS critical temperature $T_{c}^{\mathrm{BCS}}$ that was already reported in Fig.÷\ref{fig-2}(a).
This is shown in Fig.÷\ref{fig-13}, where the coupling dependence of $T^{*}$ obtained in this way is compared with that of the mean-field critical temperature 
$T_{c}^{\mathrm{BCS}}$ obtained by solving Eqs.(\ref{BCS-gap-equation}) and (\ref{BCS-density-equation}) in the limit $\Delta \rightarrow 0$.
As a further reference, Fig.÷\ref{fig-13} also reports the coupling dependence of the critical temperature $T_{c}$ that includes the effects of pairing fluctuations within the $t$-matrix approximation (as taken from Fig.1 of Ref.\cite{PPPS-2004}).

It is worth commenting that, in the literature of the BCS-BEC crossover, the mean-field critical temperature $T_{c}^{\mathrm{BCS}}$ has generically represented a pair-breaking temperature below which pre-formed pairs are formed, in such a way that the effects of precursor pairing manifest themselves between $T_{c}$ and $T_{c}^{\mathrm{BCS}}$ \cite{Levin-2005}.
With the present analysis, this temperature acquires a more physical meaning for the building up of inter-pair correlations out of intra-pair correlations that exist well above this temperature.
Correspondingly, the emphasis given in the original BCS theory \cite{Schrieffer-1964}, about the occurrence of fermionic pair correlations rather than on the  actual existence of fermion pairs, appears here to be fully justified also as far as the crossover temperature $T^{*}$ is concerned.

\section{IV. Concluding remarks}
\label{sec:conclusions}

In this paper, we have considered the pairing correlations which build up in a Fermi gas with an attractive inter-particle interaction, not only as a function of coupling throughout the BCS-BEC crossover but also as a function of temperature, both above and below the critical temperature at which the superfluid phase sets in.
This has been done in terms of two correlation functions which focus alternatively on intra-pair correlations, that depend on the relative coordinate $\boldsymbol{\rho} = \mathbf{r} - \mathbf{r'}$ between  spin-up and spin-down fermions, or on inter-pair correlations, that depend instead on the difference between the center-of-mass coordinates $\mathbf{R} =  (\mathbf{r} + \mathbf{r'})/2$ of two pairs.
It has been shown that, quite generally, the same kind of many-body diagrammatic structure can describe both correlation functions, with the only provision of setting the external spatial variables in the diagrammatic structure in an appropriate way.
This difference results, however, in drastic changes for the characteristic lengths associated with the two above correlations functions.

We have found that intra-pair correlations decrease at a rather slow rate with increasing temperature, in such a way that they survive considerably above the critical temperature.
We have also found this rate to depend on the coupling throughout the BCS-BEC crossover, in such a way that it is slower on the BEC side with respect to the BCS side of unitarity.
We have further correlated qualitatively this finding with the experimental data recently obtained on the proximity effect in the normal phase of a high-temperature (cuprate) superconductor.

In addition, we have found that above $T_{c}$ inter-pair correlations decrease with increasing temperature at a faster rate than intra-pair correlations, leading to temperature crossing between the two behaviors.
This, in turn, has led us to identify a crossover temperature $T^{*}$, such that at temperatures smaller than $T^{*}$ there is a growing importance of the inter-pair correlations which emerge out of the 
existing intra-pair correlations.
Since from a many-body point of view the occurrence of pairing correlations is a much better defined concept than the wave function of Cooper (or pre-formed) pairs, the crossover temperature $T^{*}$ identified in this way has a more sound physical basis than the pair-breaking temperature discussed thus far in the literature.

The numerical calculations were done at the level of the (non-self-consistent) $t$-matrix approximation, both below and above $T_{c}$, aiming primarily at including the effects of pairing fluctuations over and
above mean field.
Below $T_{c}$ their inclusion is essential for inter-pair correlations, but it is also important for intra-pair correlations especially as far as the short-range behavior of the pair correlation function is concerned.
This is, in turn, related to the Tan's contact that has recently attracted much interest in the context of Fermi gases.

The $t$-matrix approximation that we have utilized in the numerical calculations includes pairing fluctuations in a minimal way, which we regard sufficient to describe the main physical effects
related to intra- and inter-pair correlations we have discussed.
In this respect, even though improved diagrammatic methods like those of Refs.\cite{PS-bd-2007,Zwerger-2007} could possibly somewhat modify our numerical results in a quantitative way, they are not expected to affect in an appreciable way the overall physical framework which we have described.

\vspace{0.1cm}

\begin{center}
\begin{small}
{\bf ACKNOWLEDGMENTS}
\end{small}
\end{center}
\vspace{-0.1cm}

We are indebted to F. Marsiglio for bringing Ref.\cite{Marsiglio-1990} to our attention.
This work was partially supported by the Italian MIUR under Contract Cofin-2009 ``Quantum gases beyond equilibrium''.
 
\vspace{-0.3cm}                                                                                                                                                                                                                                                                                                                                                                                                         
\appendix
\section{APPENDIX A: PAIR CORRELATION FUNCTION AND SUM RULE}
\label{sec:appendix-A}

In this Appendix, we consider a sum rule that the pair correlation function $g_{\uparrow \downarrow}(\boldsymbol{\rho})$ defined by the expression (\ref{definition-pair-correlation-function}) should \emph{apparently} obey.
In this context, we shall have to face a rather subtle physical point that was pointed out some time ago by Bell \cite{Bell-1963}.
Accordingly, we shall see that the process of \emph{first} selecting an approximate form for $g_{\uparrow \downarrow}(\boldsymbol{\rho})$ as a function of $\boldsymbol{\rho}$ and \emph{then} performing the integral of this quantity over $\boldsymbol{\rho}$ yields a different result than doing the opposite, that is to say, choosing an approximate form directly for the integrated quantity (albeit apparently through the same kind of approximation scheme).
As pointed out by Bell, this \emph{non commutativity} of the results reflects the fact that the fluctuations of the particle number are evaluated in the grand canonical or canonical ensembles, and becomes irrelevant in the high-temperature limit when classical physics takes over.

It was mostly for this reason that in subsection II-C we have commented that, since no conservation law corresponds to the pair correlation function (\ref{definition-pair-correlation-function}), considerations about ``conserving''  diagrammatic approximations in the sense of Baym and Kadanoff \cite{BK-1961,Baym-1962} do not directly apply to it.
As a consequence, in subsection II-C the series of ladder diagrams for the T-matrix below $T_{c}$ depicted in Fig.÷\ref{fig-3}(a) was introduced for the pair correlation function mainly to recover the expected values of the Tan's contact.
In subsection II-D this series was used also above $T_{c}$ because the corresponding series of ``maximally crossed diagrams'' represents the minimal ingredient to get meaningful results for the pair correlation function (and further gives the expected result in the high-temperature limit where the non-self-consistent $t$-matrix approximation is known to become exact \cite{Combescot-2006}).

Quite generally, the sum rule that the pair correlation function $g_{\uparrow \downarrow}(\boldsymbol{\rho})$ should apparently obey can be set up as follows.
From the definition (\ref{definition-pair-correlation-function}) one gets for the volume integral of $g_{\uparrow \downarrow}(\boldsymbol{\rho})$ the expression:
\begin{equation}
\int \! d \boldsymbol{\rho} \, g_{\uparrow \downarrow}(\boldsymbol{\rho}) = \frac{1}{V} \left( \langle N_{\uparrow} N_{\downarrow} \rangle - \langle N_{\uparrow} \rangle \langle N_{\downarrow} \rangle \right)
\label{generic-volume-integral}
\end{equation}

\noindent
where $V$ is the volume occupied by the system and $N_{\sigma} = \int \! d \mathbf{r} \, \psi^{\dagger}_{\sigma}(\mathbf{r}) \psi_{\sigma}(\mathbf{r})$ is the number operator with spin $\sigma$.
On the other hand, by introducing two different chemical potentials $\mu_{\sigma}$ for each spin species, from the definition of $n_{\uparrow}$ in terms of the thermal average
\begin{equation}
n_{\uparrow} = \frac{1}{V} \, \frac{ Tr \left\{  N_{\uparrow} e^{- \beta \left( H - \mu_{\uparrow} N_{\uparrow} - \mu_{\downarrow} N_{\downarrow} \right)} \right\} }
                                                    { Tr \left\{ e^{- \beta \left( H - \mu_{\uparrow} N_{\uparrow} - \mu_{\downarrow} N_{\downarrow} \right)} \right\} }
\label{n_up-thermal-average}
\end{equation}

\noindent
where $H$ is the system Hamiltonian and $\beta = (k_{B} T)^{-1}$ the inverse temperature, one also obtains:
\begin{equation}
\left. \frac{\partial n_{\uparrow}}{\partial \mu_{\downarrow}} \right|_{T,V} = \frac{\beta}{V} \, \left( \langle N_{\uparrow} N_{\downarrow} \rangle - \langle N_{\uparrow} \rangle \langle N_{\downarrow} \rangle \right).
\label{generic-partial-up-down}
\end{equation}

\noindent
Comparison of the expressions (\ref{generic-volume-integral}) and (\ref{generic-partial-up-down}) then yields:
\begin{equation}
\int \! d \boldsymbol{\rho} \, g_{\uparrow \downarrow}(\boldsymbol{\rho}) = \frac{1}{\beta} \, \left. \frac{\partial n_{\uparrow}}{\partial \mu_{\downarrow}} \right|_{T,V} 
\label{sum-rule-pair-correlation-function}
\end{equation}

\noindent
where the limit $n_{\uparrow} \rightarrow n_{\downarrow} \rightarrow n/2$ of balanced spin populations is here understood like the rest of the paper.

The contradiction pointed out by Bell \cite{Bell-1963} is now apparent.
While the right-hand side of Eq.(\ref{sum-rule-pair-correlation-function}) is expected to vanish at zero temperature owing to the presence of the factor $\beta^{-1}$ in front of the finite value of 
$\left. \frac{\partial n_{\uparrow}}{\partial \mu_{\downarrow}} \right|_{T,V}$, the left-hand side of Eq.(\ref{sum-rule-pair-correlation-function}) is bound to remain finite once any reasonable choice of $g_{\uparrow \downarrow}(\boldsymbol{\rho})$ made beforehand is integrated over $\boldsymbol{\rho}$. 
In addition, owing to Eq.(\ref{generic-volume-integral}) the vanishing of the right-hand side of Eq.(\ref{sum-rule-pair-correlation-function}) would also imply a complete suppression of particle fluctuations, in the sense that $\langle N_{\uparrow} N_{\downarrow} \rangle = \langle N_{\uparrow} \rangle \langle N_{\downarrow} \rangle$.

Consistently with Bell's analysis, we shall here show that the ``sum rule'' (\ref{sum-rule-pair-correlation-function}) is obeyed by a ``conserving''  diagrammatic approximations in the sense of Baym and Kadanoff \cite{BK-1961,Baym-1962}, only when \emph{this approximation is made directly on the integral} of $g_{\uparrow \downarrow}(\boldsymbol{\rho})$ and not on $g_{\uparrow \downarrow}(\boldsymbol{\rho})$ itself before performing the integration.
To this end, we shall explicitly consider the extended BCS approximation, which corresponds to the series of ladder diagrams of Fig.÷\ref{fig-3}(a) and is familiar in the context of gauge invariance for the response of a superconductor to an external electromagnetic field \cite{Schrieffer-1964}. 

In this way, we shall extend Bell's analysis to finite temperature as well, and show analytically the way the identity (\ref{sum-rule-pair-correlation-function}) is satisfied in the above sense within this approximation for any temperature below $T_{c}$.
Within this approximation we shall also provide a numerical analysis, aiming at showing to what extent the numerical integration over $\boldsymbol{\rho}$ of $g_{\uparrow \downarrow}(\boldsymbol{\rho})$ given by the expression (\ref{pair-correlation-function-below-Tc}) (with the values of $\Delta$ and $\mu$ taken at the BCS mean-field level) differs from the right-hand side of Eq.(\ref{sum-rule-pair-correlation-function}) calculated also at the same level, as a function of coupling and temperature below $T_{c}$.
Finally, we shall show that the sum rule (\ref{sum-rule-pair-correlation-function}) becomes eventually satisfied at high-enough temperatures above $T_{c}$, when the expression 
(\ref{pair-correlation-function-above-Tc}) for $g_{\uparrow \downarrow}(\boldsymbol{\rho})$ that holds in this limit is integrated over $\boldsymbol{\rho}$.

Quite generally, following Bell's analysis it is possible to manipulate the right-hand side of Eq.(\ref{generic-volume-integral}) by introducing an integral over the imaginary time 
$\tau$ as follows:
\begin{eqnarray}
& & \int \! d \boldsymbol{\rho} \, g_{\uparrow \downarrow}(\boldsymbol{\rho}) + V n_{\uparrow} n_{\downarrow}  = \frac{1}{V} \langle N_{\uparrow} N_{\downarrow} \rangle
\nonumber \\
& = & \frac{1}{V \, \beta} \int_{0}^{\beta} \!\! d\tau \, \langle e^{K \tau} N_{\uparrow} e^{-K \tau} N_{\downarrow} \rangle
\nonumber \\
& = & \frac{1}{V \, \beta} \int_{0}^{\beta} \!\! d\tau \, \langle T_{\tau} \!\! \left[ \left(e^{K \tau} N_{\uparrow} e^{-K \tau}\right) N_{\downarrow} \right] \rangle
\nonumber \\
& = & - \frac{1}{\beta} \! \int_{0}^{\beta} \!\!\! d\tau \!\! \int \!\! d \boldsymbol{\rho} \, 
\langle T_{\tau} \!\! \left[\Psi_{1}(\boldsymbol{\rho},\tau) \Psi_{2}(\boldsymbol{0},0^{+}) \Psi_{2}^{\dagger}(\boldsymbol{0},0) \Psi_{1}^{\dagger}(\boldsymbol{\rho},\tau^{+}) \right] \rangle 
\nonumber \\
& = & - \frac{1}{\beta} \! \int_{0}^{\beta} \!\!\! d\tau \!\! \int \!\! d \boldsymbol{\rho} \,\,  
\mathcal{G}_{2} (\boldsymbol{\rho} \tau 1,\boldsymbol{0} 0^{+} 2;\boldsymbol{\rho} \tau^{+} 1,\boldsymbol{0} 0 2)
\label{manipulations_in_integral} 
\end{eqnarray}

\noindent
where $n_{\sigma} = \langle N_{\sigma} \rangle/V$, $K= H - \mu_{\uparrow} N_{\uparrow} - \mu_{\downarrow} N_{\downarrow}$ is the grand-canonical Hamiltonian entering
Eq.(\ref{n_up-thermal-average}), and $\mathcal{G}_{2}(1,2;1',2') = \left\langle T_{\tau}[ \Psi(1) \Psi(2) \Psi^{\dagger}(2') \Psi^{\dagger}(1') ] \right\rangle$ is the two-particle Green's function
with the Nambu representation of the field operators.

The crucial point, which has enabled us to arrive at the last line of Eq.(\ref{manipulations_in_integral}), is the consideration that \emph{the operator} $N_{\uparrow}$ \emph{commutes with the Hamiltonian} $H$ while its density $\psi^{\dagger}_{\sigma}(\mathbf{r}) \psi_{\sigma}(\mathbf{r})$ does not \cite{Bell-1963}.
For this reason, it has been possible to introduce a time variable in Eq.(\ref{manipulations_in_integral}), a process which in turn establishes connections with the continuity equation and the ensuing conservation law.

At this point one can use in the last line of Eq.(\ref{manipulations_in_integral}) the representation (\ref{Bethe-Salpeter-equation}) of the Bethe-Salpeter equation for $\mathcal{G}_{2}$, thus resulting in the following expression in terms of the single-particle Green's function $\mathcal{G}$ and the many-particle T-matrix:
\begin{eqnarray}
\int \! d \boldsymbol{\rho} \, g_{\uparrow \downarrow}(\boldsymbol{\rho}) & = & \frac{1}{\beta} \, \left\{ \int \! dk \, \mathcal{G}_{12}(k)^{2} \right.
\nonumber \\
& + & \sum_{\ell_{3} \ell_{4} \ell_{5} \ell_{6}} \int \! dk \, \mathcal{G}_{1 \ell_{3}}(k) \, \mathcal{G}_{\ell_{6} 1}(k) \, T^{\ell_{3} \ell_{4}}_{\ell_{6} \ell_{5}}(q \rightarrow 0)
\nonumber \\
& \times & \left.  \int \! dk' \, \mathcal{G}_{\ell_{4} 2}(k') \, \mathcal{G}_{2 \ell_{5}}(k') \right\}
\label{generic-expression-for-integral-of-g_updown}
\end{eqnarray}

\noindent
which holds in the present form for a homogenous system with a contact inter-particle interaction.

We next specify the T-matrix within the extended BCS approximation of Fig.÷\ref{fig-3}(a), in such a way that the expression 
(\ref{generic-expression-for-integral-of-g_updown}) reduces to:
\begin{eqnarray}
& & \int \! d \boldsymbol{\rho} \, g_{\uparrow \downarrow}(\boldsymbol{\rho}) = \frac{1}{\beta} \, \left\{ \int \! dk \, \mathcal{G}_{12}(k)^{2} \right.
\nonumber \\
& + & \int \! dk \, \mathcal{G}_{11}(k) \, \mathcal{G}_{12}(k) \, \int \! dk' \, \mathcal{G}_{22}(k') \, \mathcal{G}_{21}(k') 
\nonumber \\
& \times & \!\!\! \left. \left[ T_{\mathrm{I},\mathrm{I}}(q) + T_{\mathrm{I},\mathrm{II}}(q) + T_{\mathrm{II},\mathrm{I}}(q) + T_{\mathrm{II},\mathrm{II}}(q) \right]_{q \rightarrow 0} \right\} \, ,
\label{extended_BCS-expression-for-integral-of-g_updown}
\end{eqnarray}

\noindent
where with reference to the matrix elements (\ref{T-matrix-elements}) we have
\begin{eqnarray}
& & \left\{ T_{\mathrm{I},\mathrm{I}}(q) + T_{\mathrm{I},\mathrm{II}}(q) + T_{\mathrm{II},\mathrm{I}}(q) + T_{\mathrm{II},\mathrm{II}}(q) \right\}_{q \rightarrow 0}
\nonumber \\
& = & - \frac{ 2 \left( A(q \rightarrow 0) - B(q \rightarrow 0) \right)}{\left( A(q \rightarrow 0) - B(q \rightarrow 0) \right) \left( A(q \rightarrow 0) + B(q \rightarrow 0) \right)}
\nonumber \\
& = & - \, \frac{1}{ B(q=0)} = - \frac{1}{\left[ \int \! dk \, \mathcal{G}_{12}(k)^{2} \right]^{2}} \, .
\label{sum_of_matrix-elements}
\end{eqnarray}

\noindent
Note that, to obtain the last line of Eq.(\ref{sum_of_matrix-elements}), the BCS gap equation has been used in the form $A(q=0) = B(q=0)$ together with the definition (\ref{P_11_and_P_12}) of $B(q)$.

There then remains to show that the expression within braces on the right-hand side of Eq.(\ref{extended_BCS-expression-for-integral-of-g_updown}) coincides with
$\left. \frac{\partial n_{\uparrow}}{\partial \mu_{\downarrow}} \right|_{T,V}$ of Eq.(\ref{sum-rule-pair-correlation-function}), also calculated at the level of the BCS mean field.
To this end, we write:
\begin{equation}
\frac{\partial n_{\uparrow}}{\partial \mu_{\downarrow}} = \left. \frac{\partial n_{\uparrow}}{\partial \mu_{\downarrow}} \right|_{\Delta} + 
\left. \frac{\partial n_{\uparrow}}{\partial \Delta} \right|_{\mu_{\uparrow},\mu_{\downarrow}} \frac{\partial \Delta}{\partial \mu_{\downarrow}} \, .
\label{partial-derivatives-density}
\end{equation}

\noindent
where reference to constant values of $T$ and $V$ has been dropped for convenience.
Here, $n_{\uparrow}$ and $\Delta$ are obtained by the expressions
\begin{equation}
n_{\uparrow} = \int \! dk \, e^{i \omega_{n} \eta} \mathcal{G}_{11}(k) \,\,\, , \,\,\, \Delta = v_{0} \int \! dk  \, e^{i \omega_{n} \eta} \mathcal{G}_{12}(k)
\label{compact-BCS-density_and_gap-equations}
\end{equation}

\noindent
in terms of the BCS single-particle Green's functions corresponding to imbalanced spin populations \cite{BCS-imbalance}
\begin{eqnarray}
\left( 
\begin{array}{cc}
\mathcal{G}_{11}(k) & \mathcal{G}_{12}(k)  \\
\mathcal{G}_{21}(k) & \mathcal{G}_{22}(k)
\end{array}
\right) \!\! & = & \!\! \frac{1}{(i \omega_{n} - E_{+}(\mathbf{k})) (i \omega_{n} + E_{-}(\mathbf{k}))}
\nonumber \\
& \times &
\!\!\! \left( 
\begin{array}{cc}
i \omega_{n} + \xi_{\downarrow}(\mathbf{k})  &  - \Delta  \\
- \Delta^{*}   &  i \omega_{n} - \xi_{\uparrow}(\mathbf{k})
\end{array}
\right)
\label{G-matrix-elements-imbalanced}
\end{eqnarray}

\noindent
where now $\xi_{\sigma}(\mathbf{k}) = \mathbf{k}^{2}/(2m) - \mu_{\sigma}$, $E_{\pm}(\mathbf{k}) = E(\mathbf{k}) \pm \delta \xi(\mathbf{k})$, $E(\mathbf{k}) = \left[ \xi(\mathbf{k})^{2} + |\Delta|^{2} \right]^{1/2}$,
with $\xi(\mathbf{k}) = (\xi_{\uparrow}(\mathbf{k}) + \xi_{\downarrow}(\mathbf{k}))/2$ and $\delta \xi(\mathbf{k}) = (\xi_{\uparrow}(\mathbf{k}) - \xi_{\downarrow}(\mathbf{k}))/2$.

In this way, with reference to the first of Eqs.(\ref{compact-BCS-density_and_gap-equations}) we obtain:
\begin{eqnarray}
\left. \frac{\partial \mathcal{G}_{11}(k)}{\partial \mu_{\downarrow}} \right|_{\Delta} & = & \mathcal{G}_{12}(k)^{2} 
\noindent \\ 
\left. \frac{\partial \mathcal{G}_{11}(k)}{\partial \Delta} \right|_{\mu_{\downarrow}} & = & - 2 \, \mathcal{G}_{11}(k) \, \mathcal{G}_{12}(k) \, ,
\label{compact-density-derivatives}
\end{eqnarray}

\noindent
while with reference to the second of Eqs.(\ref{compact-BCS-density_and_gap-equations}) we obtain:
\begin{eqnarray}
\left. \frac{\partial \mathcal{G}_{12}(k)}{\partial \mu_{\downarrow}} \right|_{\Delta} & = & \mathcal{G}_{12}(k) \, \mathcal{G}_{22}(k) 
\noindent \\ 
\left. \frac{\partial \mathcal{G}_{12}(k)}{\partial \Delta} \right|_{\mu_{\downarrow}} & = & \frac{1}{\Delta} \, \mathcal{G}_{12}(k) \, - 2 \, \mathcal{G}_{12}(k)^{2}
\label{compact-gap-derivatives}
\end{eqnarray}

\begin{figure}[t]
\begin{center}
\includegraphics[angle=0,width=7.0cm]{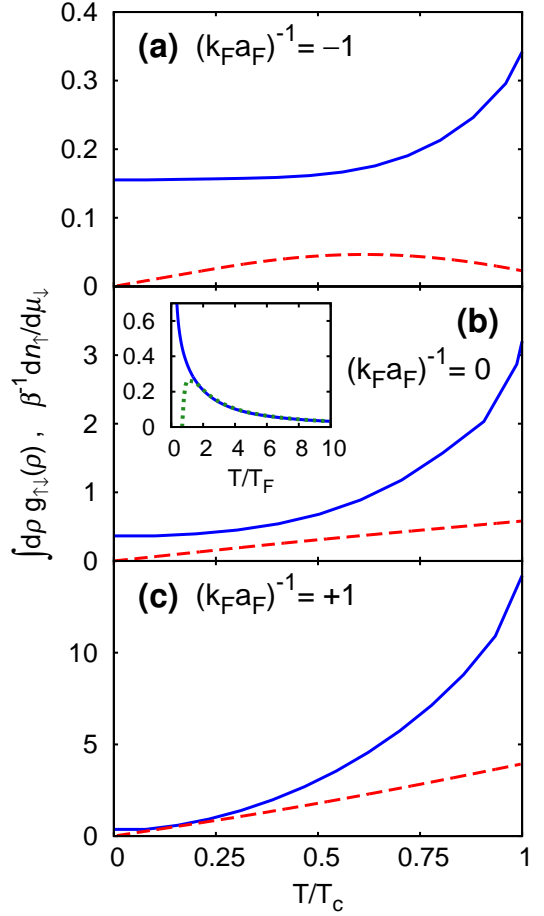}
\caption{(Color online) The temperature dependence of $\int \! d \boldsymbol{\rho} \, g_{\uparrow \downarrow}(\boldsymbol{\rho})$ below $T_{c}$ with $g_{\uparrow \downarrow}(\boldsymbol{\rho})$ 
              given by the extended BCS approximation (\ref{pair-correlation-function-below-Tc}) (full lines) is compared with the temperature dependence of 
              $\frac{1}{\beta} \, \left. \frac{\partial n_{\uparrow}}{\partial \mu_{\downarrow}} \right|_{T,V}$  given by the BCS expression (\ref{lhs-compressibility-BCS}) (dashed lines), for the couplings 
              $(k_{F} a_{F})^{-1}$: (a) $-1.0$, (b) $0.0$, (c) $+1.0$. In both quantities (which are normalized to $n/2$) the values of $\Delta$ and $\mu$ are taken at the mean-filed level. 
              The inset in the central panel shows a corresponding comparison made at unitarity in the high-temperature limit.}
\label{fig-14}
\end{center}
\end{figure}

\noindent
This yields for the derivative of $\Delta$ in Eq.(\ref{partial-derivatives-density})
\begin{equation}
\frac{\partial \Delta}{\partial \mu_{\downarrow}} = \frac{1}{2} \, \frac{\int \! dk \, \mathcal{G}_{12}(k) \, \mathcal{G}_{22}(k)}{\int \! dk \, \mathcal{G}_{12}(k)^{2}} \, ,
\label{derivative-of-gap}
\end{equation}

\noindent
such that Eq.(\ref{partial-derivatives-density}) becomes eventually:
\begin{eqnarray}
\frac{\partial n_{\uparrow}}{\partial \mu_{\downarrow}} & = & \int \! dk \, \mathcal{G}_{12}(k)^{2} 
\nonumber \\
& - &  \frac{ \int \! dk \, \mathcal{G}_{11}(k) \, \mathcal{G}_{12}(k) \, \int \! dk' \, \mathcal{G}_{22}(k') \, \mathcal{G}_{21}(k') }{\int \! dk \, \mathcal{G}_{12}(k)^{2} } 
\label{lhs-compressibility-BCS}
\end{eqnarray}

\noindent
where the limit of balanced spin populations can be restored at the end of the calculation. 
Comparison of the right-hand side of Eq.(\ref{lhs-compressibility-BCS}) with the expression within braces on the right-hand side of 
Eq.(\ref{extended_BCS-expression-for-integral-of-g_updown}) supplemented by Eq.(\ref{sum_of_matrix-elements}) proves that the ``sum rule'' 
(\ref{sum-rule-pair-correlation-function}) is indeed satisfied within the extended BCS approximation precisely in the restricted sense that we have specified above.

In practice, to quantify the violation of the sum rule (\ref{sum-rule-pair-correlation-function}) when $g_{\uparrow \downarrow}(\boldsymbol{\rho})$ is approximated by the extended BCS approximation
(\ref{pair-correlation-function-below-Tc}) (also with $\Delta$ and $\mu$ taken at the BCS mean-field level) and then integrated numerically over $\boldsymbol{\rho}$, we present in Fig.÷\ref{fig-14} the temperature dependence of $\int \! d \boldsymbol{\rho} \, g_{\uparrow \downarrow}(\boldsymbol{\rho})$ obtained in this way from $T=0$ up to $T_{c}$ for three characteristic couplings (full lines), and compare it with the corresponding temperature dependence of $\frac{1}{\beta} \, \left. \frac{\partial n_{\uparrow}}{\partial \mu_{\downarrow}} \right|_{T,V}$ where $\left. \frac{\partial n_{\uparrow}}{\partial \mu_{\downarrow}} \right|_{T,V}$ is given by the expression (\ref{lhs-compressibility-BCS}) in the limit of balanced spin populations (dashed lines).
Deviations between these two results appear to be quite substantial.

On the other hand, the inset in the central panel of Fig.÷\ref{fig-14} shows a similar comparison made in the high-temperature regime $T \gtrsim T_{F}$, with the integral of 
$g_{\uparrow \downarrow}(\boldsymbol{\rho})$ calculated numerically from the expression (\ref{pair-correlation-function-above-Tc}) (full line) and 
$\left. \frac{\partial n_{\uparrow}}{\partial \mu_{\downarrow}} \right|_{T,V}$ taken from the results of Ref.\cite{HH-2010} obtained at unitarity in terms of a high-temperature (virial) expansion (dashed line).
As anticipated, at high temperatures when classical physics takes over, the two results are seen to coincide rather accurately with each other.
\vspace{-0.2cm}
                                                                                                                                                                                                                                                                                                                                                                                                       
\appendix
\section{APPENDIX B: ASYMPTOTIC BEHAVIOR OF $\xi_{\mathrm{pair}}$ AND $\xi_{\mathrm{phase}}$ AT HIGH TEMPERATURES}
\label{sec:appendix-B}

It was shown numerically in Fig.÷\ref{fig-8} of the main text that, in the classical limit of high temperatures and irrespective of coupling, the pair coherence length $\xi_{\mathrm{pair}}$ becomes proportional to the value of the thermal wavelength with a coefficient of the order unity.
In this Appendix, we show that this result can be also obtained analytically in terms of the expressions of subsection II-D.
In addition, from the expressions of subsection III-A we shall also obtain analytically the behavior of $\xi_{\mathrm{phase}}$ at high temperatures, which is consistent with the numerical behavior reported in 
Fig.÷\ref{fig-12} of the main text.

In the classical limit, $\mu/(k_{B}T) \rightarrow - \infty$ at fixed density, such as in the expressions of subsection II-D we may consider $|\mu|/(k_{B}T) \gg 1$.
We take further $|\mu| \gg (m a_{F}^{2})^{-1}$, a condition which is satisfied at high-enough temperatures irrespective of coupling.
Accordingly, we approximate the expression (\ref{tildeP_0}) as follows:
\begin{equation}
\tilde{\Pi}_{0}(\mathbf{k};q) \simeq \frac{1}{\xi(\mathbf{k}) + \xi(\mathbf{k}+\mathbf{q}) - i \Omega_{\nu}} \, ,
\label{tildeP_0-high_T}
\end{equation}

\noindent
and take the pair propagator of the form \cite{PS-2000}:
\begin{eqnarray}
\Gamma_{0}(\mathbf{q},\Omega_{\nu}) & \simeq & - \frac{1}{\frac{m}{4 \pi a_{F}} - \frac{m^{3/2}}{4 \pi} \sqrt{ \frac{\mathbf{q}^{2}}{4 m} - 2 \mu - i \Omega_{\nu}} }
\nonumber \\
& \simeq & \frac{4 \pi}{m^{3/2}} \, \frac{1}{ \sqrt{ \frac{\mathbf{q}^{2}}{4 m} - 2 \mu - i \Omega_{\nu} } } \, .
\label{Gamma_0-high_T}
\end{eqnarray}

\noindent
We thus obtain for the last factor on the right-hand side of Eq.(\ref{volume-integral-above-Tc}):
\begin{eqnarray}
& & \!\! \int \! \frac{d\mathbf{k}}{(2 \pi)^{3}} \, \tilde{\Pi}_{0}(\mathbf{k};q)^{2} \simeq \int \! \frac{d\mathbf{k}}{(2 \pi)^{3}} \, \frac{1}{\left[\xi(\mathbf{k}) + \xi(\mathbf{k}+\mathbf{q}) - i \Omega_{\nu}\right]^{2}}
\nonumber \\
& = & \frac{m^{2}}{2 \pi^{2}} \int_{0}^{\infty} \! dk \, \frac{k^{2}}{\left[ k^{2} +  \frac{\mathbf{q}^{2}}{4} - 2 m \mu - i m \Omega_{\nu} \right]^{2}}
\nonumber \\
& = & \frac{m^{3/2}}{8 \pi} \,  \frac{1}{ \left( \frac{\mathbf{q}^{2}}{4 m} - 2 \mu - i \Omega_{\nu} \right)^{1/2} }
\label{integrand-squared-high_T}
\end{eqnarray} 

\noindent
where the last line has been obtained by a contour integration.
Correspondingly, we obtain for the last factor on the right-hand side of Eq.(\ref{second-moment-above-Tc}):
\begin{eqnarray}
& & \!\! \int \!\!\! \frac{d\mathbf{k}}{(2 \pi)^{3}} \! \left[ \nabla_{\mathbf{k}} \tilde{\Pi}_{0}(\mathbf{k};q) \right]^{2} \! \simeq \!\!
\int \!\! \frac{d\mathbf{k}}{(2 \pi)^{3}} \, \frac{ \left( \frac{2 \mathbf{k} + \mathbf{q}}{m} \right)^{2} }{\left[\xi(\mathbf{k}) + \xi(\mathbf{k}+\mathbf{q}) - i \Omega_{\nu}\right]^{4}}
\nonumber \\
& = & \frac{2 m^{2}}{\pi^{2}} \int_{0}^{\infty} \! dk \, \frac{k^{4}}{\left[ k^{2} +  \frac{\mathbf{q}^{2}}{4} - 2 m \mu - i m \Omega_{\nu} \right]^{4}}
\nonumber \\
& = & \frac{m^{1/2}}{16 \pi} \,  \frac{1}{ \left( \frac{\mathbf{q}^{2}}{4 m} - 2 \mu - i \Omega_{\nu} \right)^{3/2} }
\label{integrand-gradient_squared-high_T}
\end{eqnarray} 

\noindent
where the last line has again been obtained by a contour integration.

The volume integral (\ref{volume-integral-above-Tc}) of $g_{\uparrow \downarrow}(\boldsymbol{\rho})$ then becomes:
\begin{eqnarray}
& & \!\! \int \! d \boldsymbol{\rho} \, g_{\uparrow \downarrow}(\boldsymbol{\rho}) \simeq \frac{1}{2} \! \int \!\! \frac{d\mathbf{q}}{(2 \pi)^{3}} \, k_{B}T \sum_{\nu} \,
\frac{e^{i \Omega_{\nu} \eta}}{ \frac{\mathbf{q}^{2}}{4 m} - 2 \mu - i \Omega_{\nu} }
\nonumber \\
& = & \frac{1}{2} \! \int \! \frac{d\mathbf{q}}{(2 \pi)^{3}} \, f_{B} \!\! \left( \frac{\mathbf{q}^{2}}{4 m} - 2 \mu \right) \! \simeq \frac{1}{2} \! \! \left( \frac{m k_{B} T}{\pi} \right)^{3/2} \!\!\! e^{\frac{2 \mu}{k_{B} T}}
\label{volume-integral-above-Tc-high_T}
\end{eqnarray}

\noindent
where in the last step the Bose function $f_{B}(\epsilon) = (\exp{(\epsilon/k_{B}T)} - 1)^{-1} \simeq e^{-\epsilon/(k_{B} T)}$ has been approximated by its high-temperature form.
Correspondingly, the second moment (\ref{second-moment-above-Tc}) becomes: 
\begin{eqnarray}
& & \!\! \int \! d \boldsymbol{\rho} \, \boldsymbol{\rho}^{2} \, g_{\uparrow \downarrow}(\boldsymbol{\rho}) \simeq \frac{1}{4 m} \!\! \int \! \frac{d\mathbf{q}}{(2 \pi)^{3}} \, k_{B}T \sum_{\nu} \,
\frac{e^{i \Omega_{\nu} \eta}}{ \left(\frac{\mathbf{q}^{2}}{4 m} - 2 \mu - i \Omega_{\nu} \right)^{2}}
\nonumber \\
& = & \frac{1}{8 m} \frac{\mathrm{d}}{\mathrm{d} \mu} \! \int \! \!\! \frac{d\mathbf{q}}{(2 \pi)^{3}} \, f_{B} \!\! \left( \frac{\mathbf{q}^{2}}{4 m} - 2 \mu \right) \! \simeq \frac{1}{4 \pi} \! \!\left( \frac{m k_{B} T}{\pi} \right)^{1/2} \!\!\! e^{\frac{2 \mu}{k_{B} T}}
\label{second-moment-above-Tc-high_T}
\end{eqnarray}

\noindent
where again in the last step the Bose function has been approximated by its high-temperature form.

The above results can eventually be inserted in the definition (\ref{xi-pair-definition}) for (the square of) $\xi_{\mathrm{pair}}$, yielding the expression:
\begin{equation}
\xi_{\mathrm{pair}}^{2} = \frac{\int \! d \boldsymbol{\rho} \, \boldsymbol{\rho}^{2} \, g_{\uparrow \downarrow}(\boldsymbol{\rho})}
                                               {\int \! d \boldsymbol{\rho} \, g_{\uparrow \downarrow}(\boldsymbol{\rho})} 
\simeq \frac{\frac{1}{4 \pi} \! \!\left( \frac{m k_{B} T}{\pi} \right)^{1/2} \!\!\! e^{\frac{2 \mu}{k_{B} T}}}{\frac{1}{2} \! \! \left( \frac{m k_{B} T}{\pi} \right)^{3/2} \!\!\! e^{\frac{2 \mu}{k_{B} T}}} \, = \, \frac{1}{2m k_{B} T}                                          
\label{xi-pair-high_T}
\end{equation}

\noindent
which is valid in the high-temperature limit.
This yields $\xi_{\mathrm{pair}} \simeq (2 m k_{B} T)^{-1/2} = \lambda_{\mathrm{T}}/\sqrt{4 \pi}$, using a standard definition \cite{Huang-1963} of the \emph{thermal wavelength} 
$\lambda_{\mathrm{T}} = \sqrt{ \frac{2 \pi}{m k_{B} T} }$ (with $\hbar = 1$).
Note that, if we would have instead introduced the length $\tilde{\lambda}_{\mathrm{T}}$ such that $k_{B} T= (2 m \tilde{\lambda}_{\mathrm{T}}^{2})^{-1}$, the result (\ref{xi-pair-high_T})
would simply read $\xi_{\mathrm{pair}} \simeq \tilde{\lambda}_{\mathrm{T}}$ with a unit coefficient.

The leading behavior (\ref{Gamma_0-high_T}) can further be used to determine $\xi_{\mathrm{phase}}$ in the high-temperature limit.
Accordingly, we obtain approximately
\begin{equation}
a \simeq \frac{(2 m)^{3/2}}{8 \pi} \, \sqrt{|\mu|} \,\,\,\,\,\,\, , \,\,\,\,\,\,\, b \simeq \frac{(2 m)^{1/2}}{64 \pi} \, \frac{1}{\sqrt{|\mu|}}
\label{approximate-a_b}
\end{equation}

\noindent
for the coefficients of the expansion (\ref{small-Q-expansion}), such that
\begin{equation}
\xi_{\mathrm{phase}} \simeq \sqrt{\frac{9}{32 \, m \, |\mu|}} \, .
\label{approximate-xi_phase-preliminary}
\end{equation}

\noindent
Making use at this point of a standard expression for the chemical potential of an ideal Fermi gas valid at high temperatures ($T \gtrsim T_{F}$) \cite{Huang-1963}, the result
(\ref{approximate-xi_phase-preliminary}) becomes:
\begin{equation}
\xi_{\mathrm{phase}} \simeq \frac{3}{4} \, \frac{\xi_{\mathrm{pair}}}{\sqrt{\ln \left[ \frac{6 \, \pi^{2}}{(k_{F} \, \lambda_{T})^{3}} \right]}}
\label{approximate-xi_phase-final}
\end{equation}

\noindent
where $k_{F} \, \lambda_{T} \ll 1$ in this limit.
\vspace{-0.2cm}
                                                                                                                                                                                                                                                                                                                                                                                                       
\appendix
\section{APPENDIX C: RELATIONSHIP BETWEEN $\xi_{\mathrm{N}}$ AND $\xi_{\mathrm{pair}}$ OR $\xi_{\mathrm{phase}}$}
\label{sec:appendix-C}

In subsection II-C the experimental results of Ref.\cite{KK-2013}, about the temperature dependence of the normal coherence length $\xi_{\mathrm{N}}$, were related to our results about the temperature 
dependence of $\xi_{\mathrm{pair}}$ in the normal phase above $T_{c}$ for various couplings (cf. Fig.÷\ref{fig-9}(a) of the main text).
In this Appendix, we substantiate our argument for having associated $\xi_{\mathrm{N}}$ with $\xi_{\mathrm{pair}}$ and not with $\xi_{\mathrm{phase}}$ in the context of the experiment of 
Ref.\cite{KK-2013}, where the temperature window is comprised between about $1.5 \, T_{c}$ and $3.0 \, T_{c}$ and therefore \emph{is not too close to} $T_{c}$.

\begin{figure}[h]
\begin{center}
\includegraphics[angle=0,width=8.2cm]{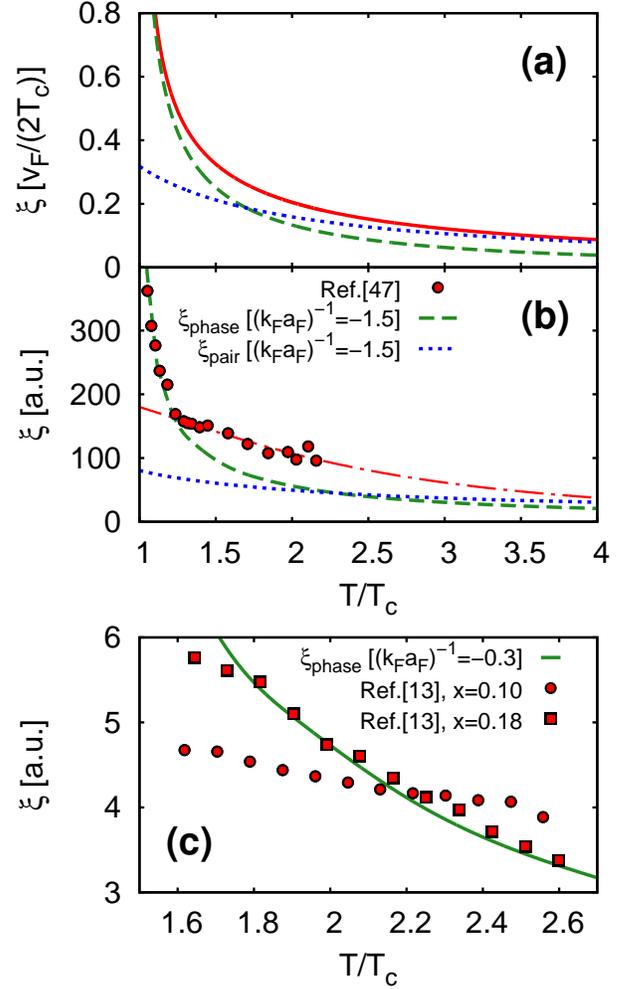}
\caption{(Color online) (a) Temperature dependence of the length scale $\xi$ over which superconducting correlations survive in the normal phase according to Ref.\cite{Kogan-1982} (full line).
                                     Also shown are the curves from Ref.\cite{Kogan-1982} that can be identified with $\xi_{\mathrm{phase}}$ (dashed line) and $\xi_{\mathrm{pair}}$ (dotted line) of the present approach.
                                     Here $v_{F}$ is the Fermi velocity.
                                     (b) The data from Fig.4 of \cite{Deutscher-1991} (circles) are compared with the temperature dependence of $\xi_{\mathrm{phase}}$ (dashed line) and $\xi_{\mathrm{pair}}$ (dotted 
                                     line) calculated for the coupling $(k_{F} a_{F})^{-1} = -1.5$.
                                     The dash-dotted line extrapolates high-temperature behavior of the data.
                                     (c) Attempts to fit the data from Ref.\cite{KK-2013} in terms of $\xi_{\mathrm{phase}}$, instead of $\xi_{\mathrm{pair}}$ as in Fig.÷\ref{fig-9}(a) of the main text.
                                     All attempts have failed for the for the under-doped material (circles).}
\label{fig-15}
\end{center}
\end{figure}

It was shown some time ago by Kogan \cite{Kogan-1982} that the appropriate coherence length of a normal metal in a proximity system changes with temperature in a continuous (albeit nontrivial) fashion, 
from what we have here identified with $\xi_{\mathrm{phase}}$ close to $T_{c}$, to what we have identified with $\xi_{\mathrm{pair}}$ somewhat above $T_{c}$ \cite{footnote-Kogan}.

Kogan's approach holds in what would be referred to as the \emph{extreme} BCS limit (namely, $(k_{F} a_{F})^{-1} \ll -1$) in the language of the BCS-BEC crossover.
Nevertheless, on physical grounds one expects Kogan's result (namely, $\xi_{\mathrm{phase}}$ close to $T_{c}$ turning into $\xi_{\mathrm{pair}}$ somewhat above $T_{c}$) to continue to hold even away from this limit. 
Lacking at present a more complete theory that would extend Kogan's approach throughout the BCS-BEC crossover, this was the reason why in Fig.÷\ref{fig-9}(a) of the main text we compared the experimental data of Ref.\cite{KK-2013} with the temperature dependence of $\xi_{\mathrm{pair}}$.
Our reasoning is substantiated by the plots reported in Fig.÷\ref{fig-15}.

In particular, in panel (a) of Fig.÷\ref{fig-15} the full line shows the temperature dependence of Kogan's $\xi$ (as given by Eq.(25) of Ref.\cite{Kogan-1982}), the dashed line represents the extrapolation over an extended temperature interval of its limiting (Ginzburg-Landau) behavior close to $T_{c}$ (as given by Eq.(27) of Ref.\cite{Kogan-1982}), and the dotted line represents the extrapolation down to $T_{c}$ of its limiting high-temperature behavior (as given by Eq.(26) of Ref.\cite{Kogan-1982}).
In the language of the present paper, the dashed line can thus be identified with $\xi_{\mathrm{phase}}$ and the dotted line with $\xi_{\mathrm{pair}}$ \cite{footnote-Kogan}. 
Note that the two extrapolated lines cross each other at about $1.75 \, T/T_{c}$ and that the full line lies always above these two extrapolated curves.

Analogous features appear when comparing the older data reported in Fig.4 of \cite{Deutscher-1991} (which get quite close to $T_{c}$) with our present results for $\xi_{\mathrm{phase}}$ and 
$\xi_{\mathrm{pair}}$ in the context of the BCS-BEC crossover. 
Panel (b) of Fig.÷\ref{fig-15} shows this comparison. 
Here, the data from Fig.4 of \cite{Deutscher-1991} (circles) are compared with our curves for $\xi_{\mathrm{phase}}$ (dashed line) and $\xi_{\mathrm{pair}}$ (dotted line) calculated for the common coupling value $(k_{F} a_{F})^{-1} = -1.5$ (which is still in the BCS - albeit not too extreme - regime).   
The dash-dotted line is a guide for the eye which extrapolates the high-temperature behavior of the data, with reference to which the sudden rise of the data at about $1.25 \, T/T_{c}$ is evident.
In this way $\xi_{\mathrm{phase}}$ (dashed line) and $\xi_{\mathrm{pair}}$ are seen to represent reasonably the limiting behaviors of the data.

Finally, in panel (c) of Fig.÷\ref{fig-15} we reconsider the recent data of Ref.\cite{KK-2013} (which do not get close to $T_{c}$) and try to fit them with our calculation for $\xi_{\mathrm{phase}}$, instead of 
$\xi_{\mathrm{pair}}$ as we did in Fig.÷\ref{fig-9}(a) of the main text. 
While we were ably to find a coupling value ($(k_{F} a_{F})^{-1} = -0.3$) for which $\xi_{\mathrm{phase}}$ (full line) can reasonably represent the data for the optimally-doped material (LSCO-0.18, squares),
all attempts we have made to represent with $\xi_{\mathrm{phase}}$ the data for the under-doped material (LSCO-0.10, circles) failed even though we have spanned the coupling $(k_{F} a_{F})^{-1}$ throughout the BCS-BEC crossover.




\begin{thebibliography}{99}
\bibitem{footnote-xi-phase} In this context, the symbol $\xi_{\mathrm{phase}}$ refers to the fact that in a superconductor (or in a superfluid) 
                                            it is the \emph{phase} of the order parameter to be fixed by the broken-symmetry.
\bibitem{footnote-1} For large values of $\mathbf{Q}$, the integrand of Eq.(\ref{F-parallel-beyond-mean-field-final}) acquires the asymptotic form $4 \pi / (m |\mathbf{Q}|)$ irrespective
                                of coupling, such that at short distances $F_{\parallel}(\mathbf{R} - \mathbf{R}')$ behaves like $2 / (\pi m |\mathbf{R} - \mathbf{R}'|^{2})$. A similar result holds also 
                                for the expression (\ref{F-above-Tc}) above $T_{c}$.
\bibitem{footnote-Kogan}  When referring to the expressions of Ref.\cite{Kogan-1982}, we recall that $\xi$ represents the length scale over which superconducting correlations 
                                          survive in the normal phase, while $\xi_{N}$ can be identified with our $\xi_{\mathrm{pair}}$ not too far from $T_{c}$ and in the extreme BCS limit, 
                                          as seen from the expression (12) therein.
\end{thebibliography}
\end{document}